\documentclass[pdflatex,sn-mathphys-num,twocolumn]{sn-jnl}

\usepackage[numbers]{natbib} 
\usepackage{mathtools}
\usepackage{graphicx}
\usepackage{dcolumn}
\usepackage{bm}
\usepackage{physics}
\usepackage{amsmath}
\usepackage{amssymb}
\usepackage{amsthm}
\usepackage{booktabs} 
\usepackage{mathrsfs}
\usepackage{makecell}
\usepackage{braket}
\usepackage{xcolor}
\usepackage[roman]{complexity}
\usepackage{tikz}
\usetikzlibrary{quantikz2}
\usepackage{pdfpages}
\usepackage{pgffor}
\usepackage{caption}
\usepackage{subcaption}
\usepackage{multirow}%
\usepackage{amsmath,amssymb,amsfonts}%
\usepackage{amsthm}%
\usepackage{mathrsfs}%
\usepackage[title]{appendix}%
\usepackage{xcolor}%
\definecolor{mygray}{gray}{0.8} 
\usepackage{textcomp}%
\usepackage{manyfoot}%
\usepackage{booktabs}%
\usepackage{algorithm}%
\usepackage{algpseudocode}%
\usepackage{listings}%
\usepackage{geometry}
\usepackage{lmodern} 
\usepackage{mathabx}
\usepackage{wasysym}
\usepackage{graphicx} 
\usepackage{subcaption} 
\usepackage {pythonhighlight}
\theoremstyle{thmstyleone}%
\theoremstyle{thmstyletwo}%

\theoremstyle{thmstylethree}%

\begin{document}

\title[Article Title]{The Efficiency Frontier: Classical Shadows versus Direct Quantum Measurement}

\author[1]{\fnm{Shuowei} \sur{Ma}}
\author[1]{\fnm{Junyu} \sur{Liu}}

\affil[1]{\orgdiv{Department of Computer Science}, \orgname{The University of Pittsburgh}, \orgaddress{\city{Pittsburgh}, \postcode{PA 15260}, \country{USA}}}

\maketitle
\textbf{Interfacing quantum and classical processors is an important subroutine in full-stack quantum algorithms. The so-called ``classical shadow'' method efficiently extracts essential classical information from quantum states, enabling the prediction of many properties of a quantum system from only a few measurements. However, for a small number of highly non-local observables, or when classical post-processing power is limited, the classical shadow method is not always the most efficient choice. Here, we address this issue quantitatively by performing a full-stack resource analysis that compares classical shadows with direct quantum measurement. Under certain assumptions, our analysis illustrates an efficiency frontier between classical shadows and direct quantum measurement in the information-extraction stage. For observables expressed as linear combinations of Pauli matrices, the classical shadow method outperforms direct measurement when the number of observables is large and the Pauli weight is small. For observables in the form of large Hermitian sparse matrices, the classical shadow method shows an advantage when the number of observables, the sparsity of the matrix, and the number of qubits fall within a certain range. The key parameters influencing this behavior include the number of qubits $n$, observables $M$, sparsity $k$, Pauli weight $w$, accuracy requirement $\epsilon$, and failure tolerance $\delta$. We also compare the resource consumption of the two methods on different types of quantum computers and identify break-even points where the classical shadow method becomes more efficient, which vary depending on the hardware. This paper opens a new avenue for quantitatively designing optimal strategies for hybrid quantum-classical tomography and provides practical insights for selecting the most suitable quantum measurement approach in real-world applications.}

\section{Introduction}\label{sec1}
\hspace{1em} Quantum computing is one of the most important areas driving the development of next-generation computing technologies \cite{nielsen2010quantum}. Quantum computing leverages a much larger set of quantum states compared to classical computing \cite{daley2022practical}, offering the potential to significantly enhance computational speed and to solve certain problems that classical computing cannot address efficiently \cite{kim2023scalable, o2023purification, kandala2019error, cai2023quantum}. This is specifically manifested in search \cite{grover1996fast}, factorization \cite{shor1999polynomial}, and simulation \cite{farhi2014quantum}. However, in many quantum algorithms such as the Harrow-Hassidim-Lloyd (HHL) algorithm \cite{Harrow2008QuantumAF}, downloading the information of quantum states and extracting information from quantum states through measurement is an essential part of quantum computation. Therefore, downloading has become a major challenge in quantum computation. If a great deal of time is spent on downloading, it will reduce the quantum advantage. Accordingly, efficient tomography of quantum states has emerged as a significant research area. The recently proposed classical shadow method \cite{huang2020predicting} harnesses both classical and quantum capabilities, allowing for the prediction of various quantum system properties with a logarithmic number of measurements, is now regarded as one of the mainstream subroutines in quantum information processing. Moreover, the classical shadow method has found widespread use in both experimental and theoretical computer science research \cite{struchalin2021experimental, zhang2021experimental}.

Despite its theoretical advantages, such as logarithmic scaling on number of copies of quantum states required, the practical benefits of using classical shadows compared to direct quantum measurement strategies in specific experimental setups remain unclear. The efficiency and applicability of the classical shadow approach depend significantly on the type of observables being predicted. Even though the classical shadow method is highly efficient in estimating local observables (e.g., single-qubit expectation values), its practical benefit has been limited when dealing with global or complex observables (e.g., non-local correlation functions or complicated Hamiltonians) and when the number of observables is relatively small. Additionally, shadow-based optimizations sometimes require additional post-processing time, making it challenging to assess its general effectiveness compared to direct quantum measurement strategies, especially for problems that involve higher-order correlations.

In this work, we address these challenges by evaluating the resource requirements of the classical shadow method using a quantitative approach. To determine the parameter regimes in which the classical shadow method provides practical benefits, we conduct an extensive cost-based analysis. Since the types of observables vary significantly depending on the scenario, we define two representative types of observables: one is a linear combination of Pauli matrices with Pauli weight $w$ and precision $\varepsilon$, and the other is a Hermitian matrix of dimension $2^n \times 2^n$ (where $n$ is the number of qubits) with sparsity $k$ and precision $\varepsilon$, defined as the maximum number of non-zero elements per row. In our result, we derive analytical expressions to estimate the resource consumption of both approaches.

\begin{figure*}[htbp]
    \centering 
    \includegraphics[width=1\textwidth, trim=20pt 100pt 80pt 135pt, clip]{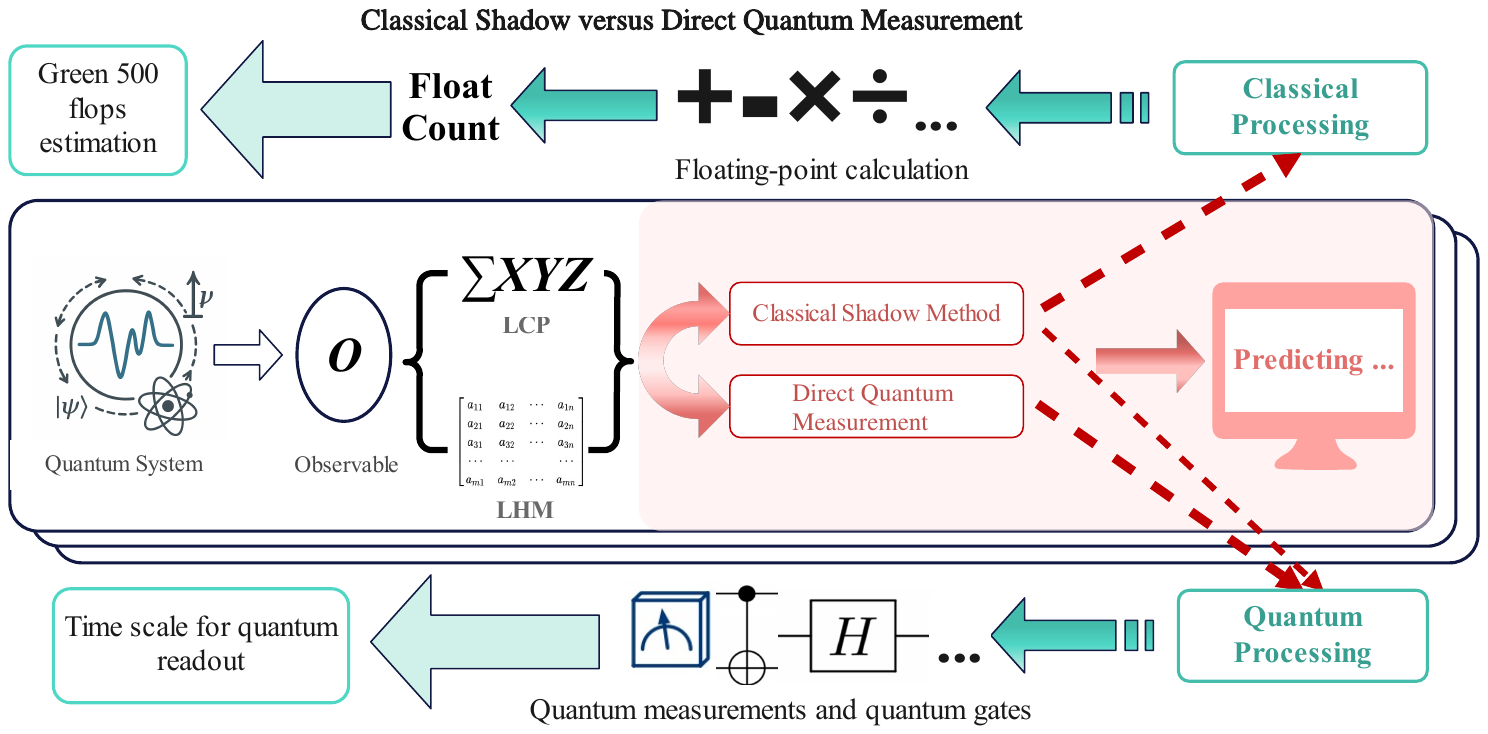}
    \captionsetup{margin={0cm,0cm}} 
    \caption{\textbf{Overall illustration of results in our paper}}  
    \label{fig:prompt} 
\end{figure*}

Since the classical shadow method is inherently a hybrid algorithm that involves both quantum and classical computation, the quantum part includes random classical rotations of the quantum state and quantum measurements, followed by transmitting the results to a classical computer. The classical part then applies an inverse transformation and performs averaging using the median-of-means algorithm in a quantum-robust manner \cite{huang2020predicting}. We estimate the resource consumption of the quantum and classical components separately. For the classical part, the cost is measured in terms of floating-point operations, a widely accepted metric in high-performance computing. Given that the classical shadow method is particularly advantageous when applied to large-scale quantum systems, ordinary desktop or laptop computers may be insufficient for the classical post-processing. Therefore, we utilize data from supercomputing clusters, referring to performance data from the TOP500 list \cite{top500_2024}.

The quantum component primarily involves quantum gates and measurements, whose time scales on basic operations have been estimated in \cite{AbuGhanem_2025}. Our optimization goal is to minimize total resource cost under a constraint on the error rate, which we set to approximately $0.01$, a value we treat as comparable to the target precision $\varepsilon$. Finally, we compare the classical shadow method with direct quantum measurement method across both the input level and the full computational stack. Direct quantum measurement can be seen as a fully quantum approach and similarly employs probabilistic bounds, such as Hoeffding's inequality \cite{hoeffding1994probability} and Chebyshev's inequality \cite{chebyshev1867valeurs}, to estimate statistical error. We conduct a comprehensive resource analysis of direct quantum measurement as well, including accurate estimations of the number of measurements required.

We organize our work as follows: In Section~\ref{sec2}, we compare the resource usage of the classical shadow technique and the direct quantum measurement method under various parameter settings. Section~\ref{sec3} explains how these results are obtained, including numerical simulations and theoretical analysis. Section~\ref{sec4} presents conclusions, suggests directions for future work, and includes additional technical details.

\section{Results}\label{sec2}
\hspace{1em}This chapter presents the main theoretical and numerical findings of this study. We compare two observable input schemes based on the classical shadow method and their corresponding direct quantum measurement strategies. For clarity, we introduce abbreviations for the two input schemes: the observable input scheme using linear combinations of Pauli matrices is referred to as LCP, while the scheme using large Hermitian matrices is referred to as LHM. 
\subsection{Linear Combination of Pauli Matrices (LCP)}
\textbf{Statement 1.}
\textit{Given \( M \) observables \( \{O_i\} \) (where \( 1 \leq i \leq M \)) in the form of a linear combination of tensor products of Pauli matrices, each observable consists of \( L \) terms in the summation, with each term having a Pauli weight of \( w \), and the coefficients following a standard normal distribution. Then, consider a quantum system with a ground state density matrix \( \rho \) (unknown) and \( n \) qubits. The classical shadow method predicts the expectations \( \hat{o}_i \) of the \( M \) observables \( \{O_i\} \) (where \( 1 \leq i \leq M \)), such that
\[
|\hat{o}_i - \text{tr}(O_i \rho)| \leq \varepsilon \text{ for all } 1 \leq i \leq M
\]
holds with a probability of at least \( 1 - \delta \).\\
\\
(a) The quantum resources required by classical shadow method are:}
\begin{align}
G &\lesssim n \cdot T \\
T &\lesssim \frac{17L \cdot 3^{w}}{\varepsilon^{2}} \cdot \log\left(\frac{2M}{\delta}\right)
\end{align}

\textit{Here, \( G \) represents the number of quantum gates needed, and \( T \) represents the number of measurements required.}\\
\\
\textit{(b) The classical resource required by classical shadow method, i.e., the number of floating-point operations, is \( C \) FLOPs, where,}
\begin{align}
C &\lesssim M \cdot L \cdot \left( T \cdot \left( \frac{1}{3} \right)^w \cdot (w + 1) + 2 \cdot \log\left( \frac{2M}{\delta} \right) + 2 \right) \tag{3}
\end{align}
\textit{At this time, \( T \) also satisfies Equation} (2)\\
\\
This statement can be regarded as a rigorous, reasonable, and saturated estimation of resource usage, supported by both mathematical proofs and numerical experiments. The detailed descriptions of the proofs and mathematical experiments are provided in the Appendix~\ref{sec:b1}. In comparison with the classical shadow technique, where the measurement count is reduced by a factor of $O(\log(M))$ in the original formulation, Statement~1 makes it more convenient to estimate resource usage.

The same applies to the estimation of expectation values of observables in the form of linear combinations of Pauli matrices using direct quantum measurement. In this case, similar statements hold, but the resource estimation focuses solely on the number of quantum measurement operations. The details regarding the proof of Statement 2 and the corresponding mathematical experiments are included in the Appendix~\ref{sec:b2}.\\
\\
\textbf{Statement 2.}\textit{
For the estimation of the expectation values of observables, \( T' \) represents the number of measurements required by the direct quantum measurement} method, where,
\begin{align}
T' &\lesssim \frac{0.5ML^3 }{\epsilon^2} \log\left( \frac{2ML}{\delta} \right)
\end{align}
\textit{The meanings of these parameters are the same as those defined in Statement 1.}\\
\\
For the input method of observables involving Large Hermitian Matrices (LHM), we perform resource estimation and provide corresponding statements analogous to those presented for the LCP. Analyses and statements that are identical to those for the LCP will not be repeated here. Similarly, we present two relevant statements. The same notation is used as follows for consistency.
\subsection{Large Hermitian Matrices (LHM)}
\begin{figure*}
    \centering
\begin{subfigure}{1.12\textwidth}
    \centering
    \includegraphics[width=\textwidth, trim=90pt 600pt 20pt 70pt, clip]{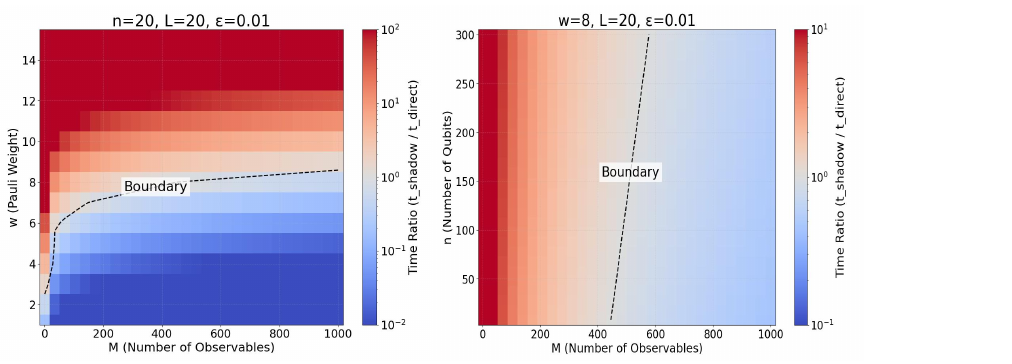}
    \captionsetup{margin={0cm,3cm}} 
    \caption{Runtime Ratio (LCP): $\frac{\text{Classical Shadow Overhead}}{\text{Direct quantum measurement Overhead} }$}
    \label{subfi:a}
\end{subfigure}
    \begin{subfigure}{1.07\textwidth}
        \centering
        \includegraphics[width=\textwidth, trim=90pt 600pt 50pt 70pt, clip]{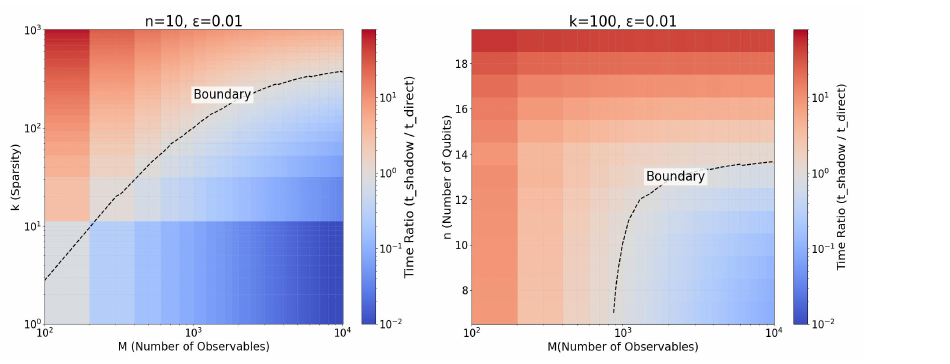}
         \captionsetup{margin={0cm,3cm}} 
        \caption{Runtime Ratio (LHM): $\frac{\text{Classical Shadow Overhead}}{\text{Direct quantum measurement Overhead} }$}
        \label{subfi:b}
    \end{subfigure}    
    \begin{subfigure}{0.48\textwidth}
        \centering
        \includegraphics[width=\textwidth, trim=10pt 10pt 10pt 0pt, clip]{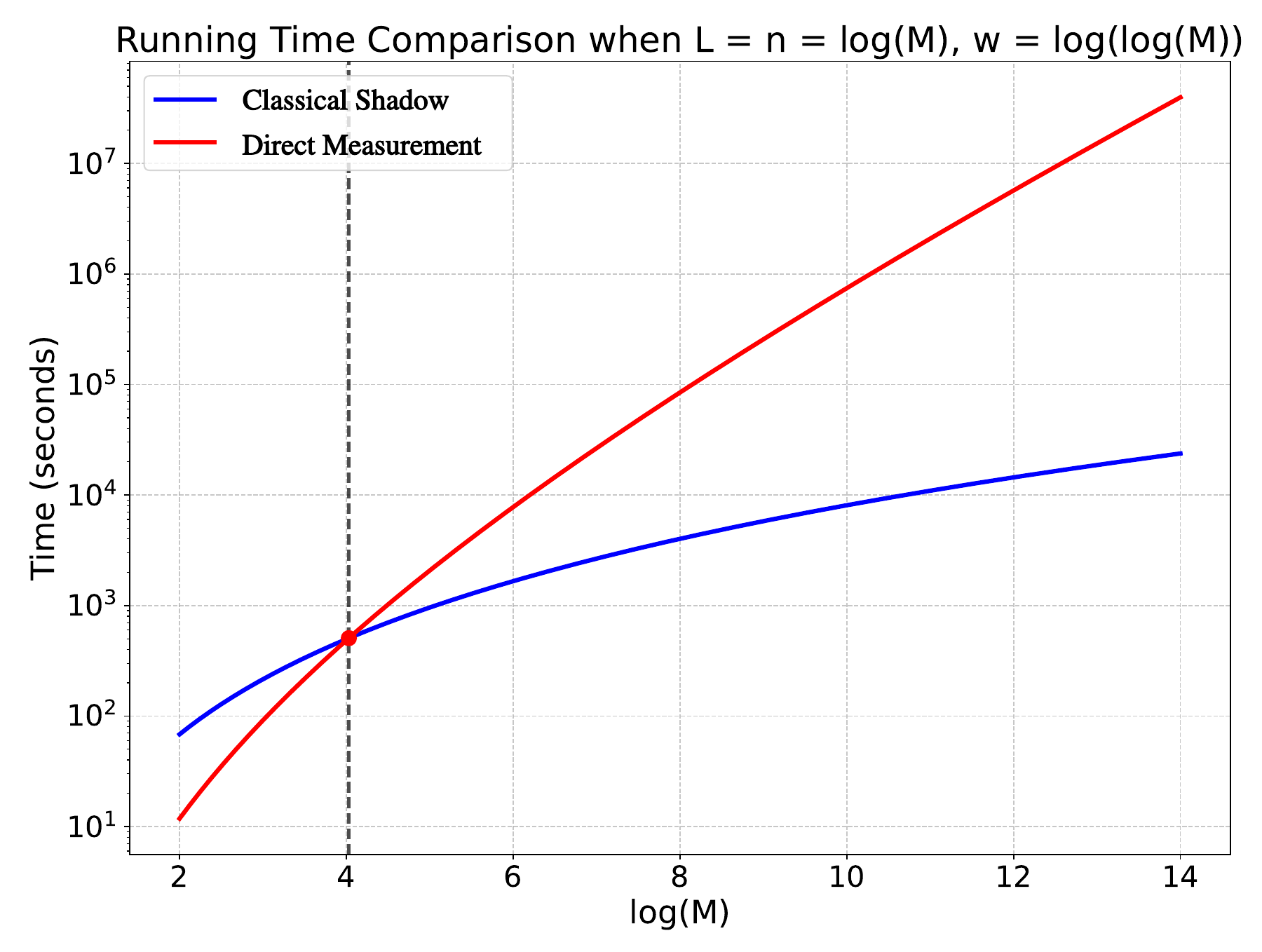}
        \caption{Runtime Comparison (LCP).}
        \label{subfi:c}
    \end{subfigure}
    \begin{subfigure}{0.48\textwidth}
        \centering
        \includegraphics[width=\textwidth, trim=10pt 10pt 10pt 0pt, clip]{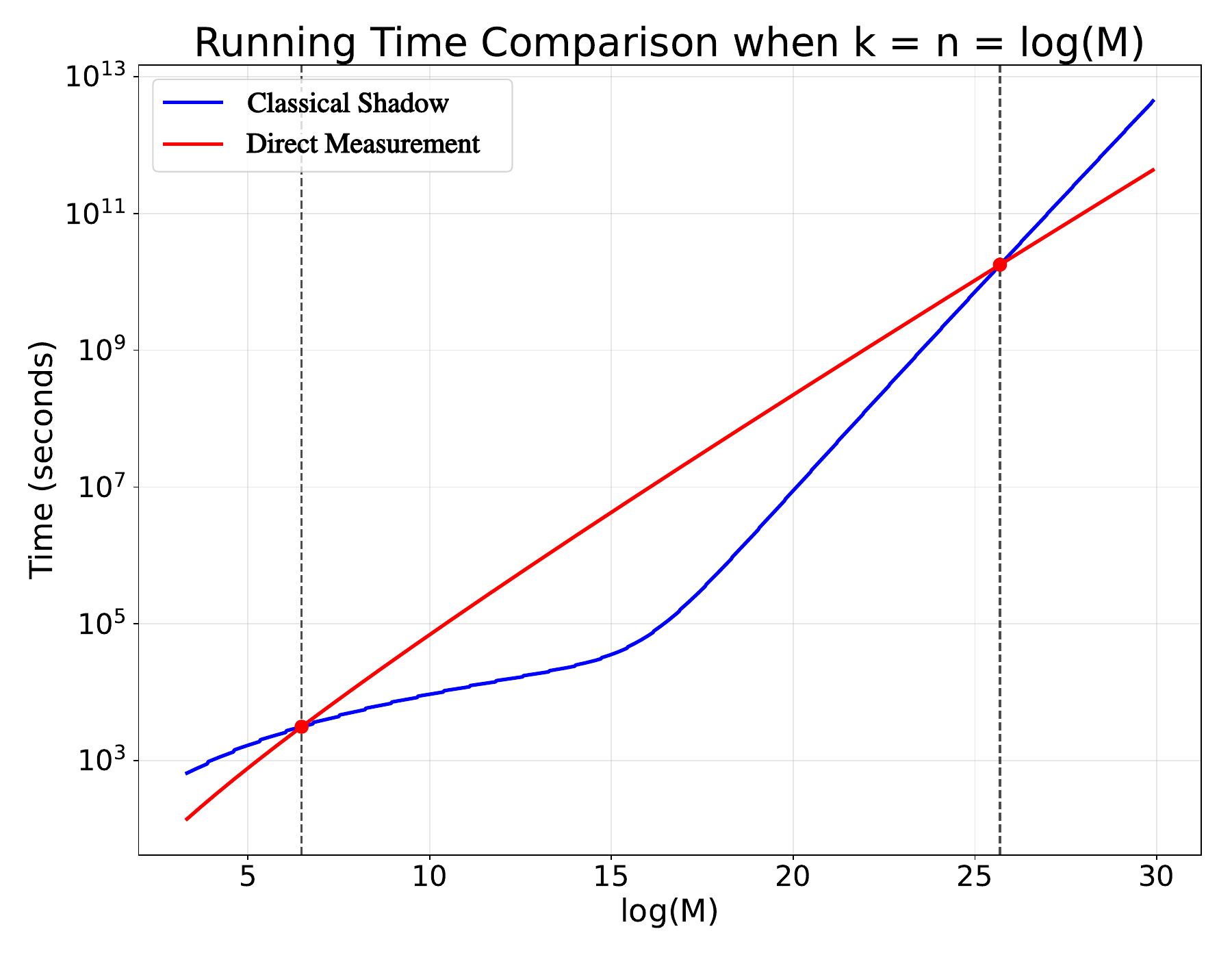}
        \caption{Runtime Comparison (LHM).}
        \label{subfi:d}
    \end{subfigure}

        \caption{
    \textbf{Comparison of Classical Shadow Method and Direct Quantum Measurement Method.} In the plots, $M$ represents the number of observables, $n$ is the number of qubits, $w$ is Pauli weight, $L$ is the number of terms per observables, $k$ is the sparsity,  $\epsilon$ is the precision tolerance of the prediction.
    (a) \& (b): Heatmaps demonstrating the comparison between the classical shadow method and the direct quantum measurement method for LCP observables and LHM observables under different problem parameters. Blue region indicates that the classical shadow method is faster than the direct quantum measurement method, while red region indicates the opposite.
    (c) \& (d): Runtime comparison between classical shadow method and direct quantum measurement method for LCP observables and LHM observables. LCP: assuming number of terms per observables and number of qubits as $\log (M)$, Pauli weight as $\log \log(M)$. LHM: number of qubits and sparsity as $\log (M)$.
    }
    \label{fig:big_figure}
\end{figure*}

\begin{figure*}
    \centering
    \begin{subfigure}{0.49\textwidth}
        \centering
        \includegraphics[width=\textwidth, trim=10pt 0pt 10pt 0pt, clip]{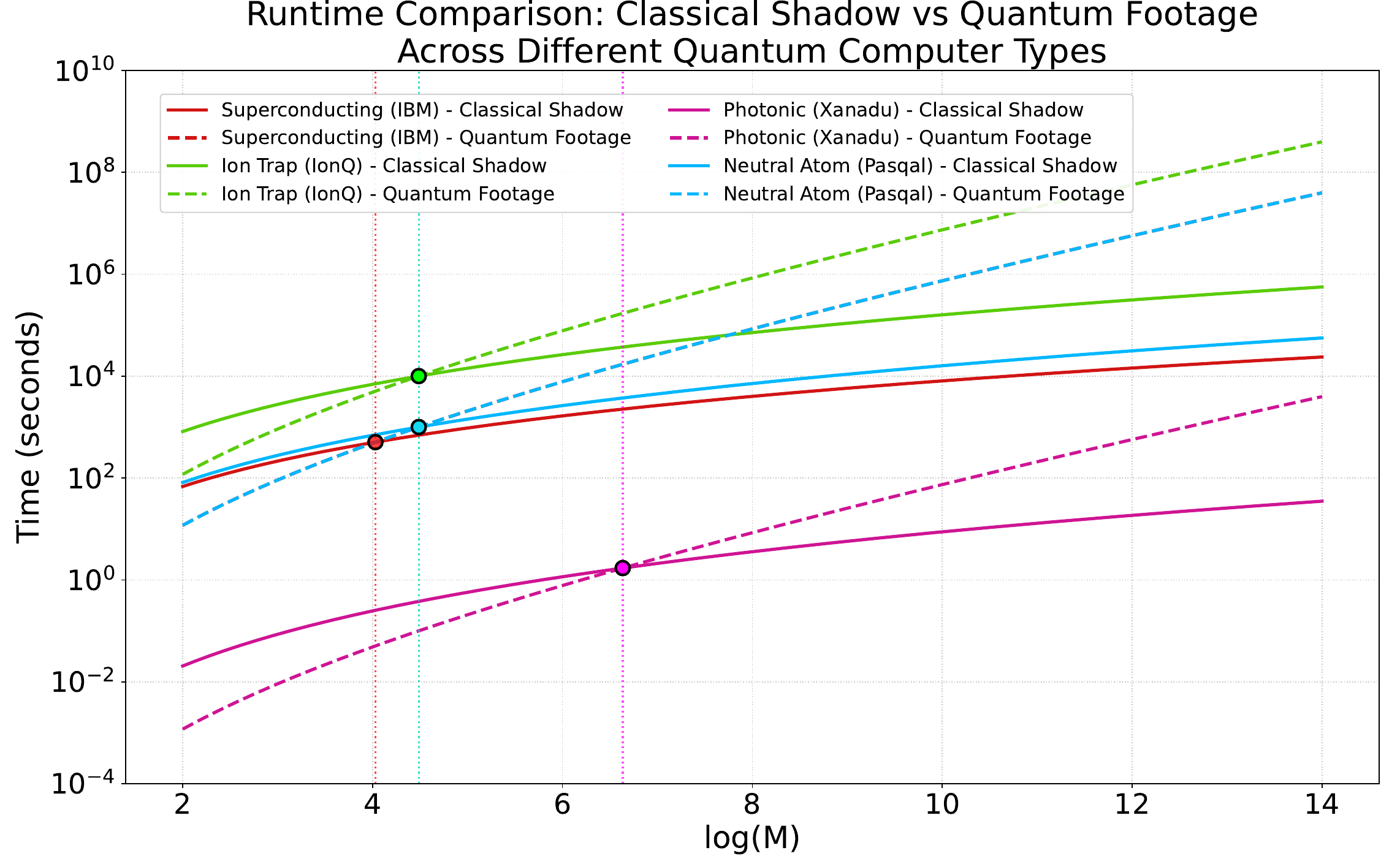}
         \captionsetup{margin={0cm,0cm}} 
        \caption{Runtime Comparison (LCP).}
        \label{subfig2:a}
    \end{subfigure}
    \begin{subfigure}{0.49\textwidth}
        \centering
        \includegraphics[width=\textwidth, trim=10pt 0pt 10pt 0pt, clip]{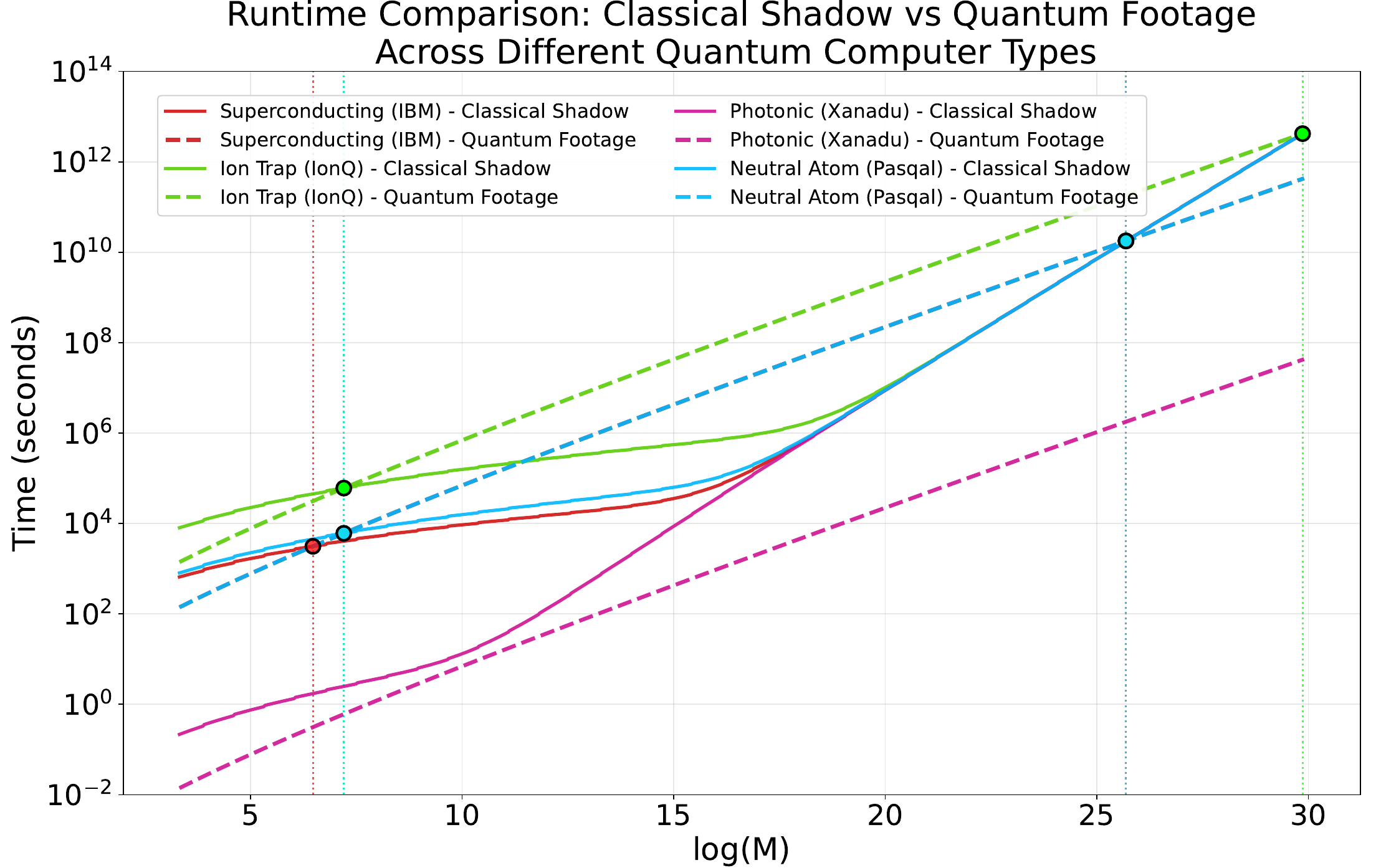}
         \captionsetup{margin={0cm,0cm}} 
        \caption{Runtime Comparison (LHM).}
         \label{subfig2:b}
    \end{subfigure}

        \caption{
    \textbf{Comparison of Classical Shadow Method and Direct Quantum Measurement Method of four types of quantum computer.} In the plots, $M$ represents the number of observables, $n$ is the number of qubits, $w$ is Pauli weight, $L$ is the number of terms per observables. $k$ is the sparsity,  $\epsilon$ is the precision tolerance of the prediction.
    (a) \& (b): 
   Runtime comparison between classical shadow method and direct quantum measurement method of four different types of quantum computer for LCP observables and LHM observables. LCP: assuming number of terms per observables and number of qubits as $\log (M)$, Pauli weight as $\log \log(M)$. LHM: number of qubits and sparsity as $\log (M)$.
    }
\label{fig:big_figure2}
\end{figure*}

\textbf{Statement 3.}\textit{
Given \( M \) observables \( \{O_i\} \) (where \( 1 \leq i \leq M \)) in the form of \( 2^n \times 2^n \) Hermitian matrices, where \( n \) is the number of qubits in the given quantum system and \( k \) is the sparsity of the matrices (each row has at least \( k \) non-zero elements, each row of elements in the matrix is sampled from the standard normal distribution), the classical shadow method predicts the expectations \( \hat{o}_i \) of the \( M \) observables \( \{O_i\} \) (where \( 1 \leq i \leq M \)), such that
\[
|\hat{o}_i - \text{tr}(O_i \rho)| \leq \varepsilon \text{ for all } 1 \leq i \leq M
\]
with probability at least \( 1 - \delta \).\\
\\
(a) The quantum resources required by classical shadow method are(when \( k \ll 2^n \)):} 
\begin{align}
G& \lesssim n \cdot T \label{eq:g_relation} 
\end{align}
\begin{align}
T& \lesssim  \frac{34 \cdot \left[ k \sqrt{\frac{2}{\pi}} + \sqrt{k \left(1 - \frac{2}{\pi}\right) \cdot 2n \log 2} \right]^2 \cdot \left[ 2 \log(2M/\delta) \right]}{\varepsilon^2} \label{eq:t_relation}
\end{align}
\textit{Here, \( G \) represents the number of quantum gates needed, and \( T \) represents the number of measurements required.}\\
\\
\textit{(b) The classical resource required by classical shadow method, i.e., the number of floating-point operations, is \( C \) FLOPs, where,}
\begin{align}
C &\lesssim T \left( 8(4^n - 1) + 32n + 7M \cdot 2^n k \right)
\end{align}
\textit{At this time, \( T \) also satisfies Equation} (5)\\
\\
We also have a similar statement, which only needs to estimate the resource consumption based on the number of quantum measurements. All details of the proofs and the mathematical experiments are placed in the Appendix~\ref{sec:B}.\\
\\
\textbf{Statement 4.}\textit{
For the estimation of the expectation values of observables, \( T' \) represents the number of measurements required by the direct quantum measurement} method, where,
\begin{align}
T'&\lesssim M \cdot \frac{16k \log\left( \frac{2M}{\delta} \right)}{\epsilon^2}
\end{align}
\textit{The meanings of these parameters are the same as those defined in Statement 3.}\\
\\
\indent Lastly, we multiplex hardware parameters to highlight the relationship between specific runtime overheads and various input values. We find that the dominant computation time for each quantum gate in a superconducting quantum computer is approximately 10 nanoseconds, while the time for a single quantum measurement is on the order of 10 microseconds \cite{AbuGhanem_2025}. In comparison, the floating-point operation rate of a supercomputer is on the order of $10^{15}$ FLOP/s \cite{top500_2024}. We provide a comprehensive explanation of the estimation techniques in Section~\ref{sec3}. Figure~\ref{fig:big_figure} presents heat maps illustrating the discrepancies between the classical shadow and direct quantum measurement approaches for two types of observable inputs: LCP ($M, L, n, w$) and LHM ($M, k, n$).\\
\indent  In these heat maps, for both the LCP and LHM cases, the blue regions represent areas where the ratio of the overhead incurred by the classical shadow method to that of the direct quantum measurement method is less than or equal to one, i.e., $t_{\text{shadow}} / t_{\text{footage}} \leq 1$. This indicates that the classical shadow method incurs less overhead compared to the direct quantum measurement approach. Consequently, these regions may suggest the potential advantages of employing a hybrid classical-quantum strategy. Moreover, under certain assumptions, Figure~\ref{fig:big_figure} also presents curves showing how the runtime of the classical shadow and direct quantum measurement methods changes as various variables increase for the two observable forms. Similarly, Figure~\ref{fig:big_figure2} presents comparative curves of the runtime of the two prediction methods for the two observable forms, under certain assumptions, across multiple types of quantum computers. It should be emphasized that in our framework all parameters are configurable, and the assumptions can be adjusted with respect to future quantum computing architectures.
\\
\indent As a summary, we have the following primary conclusions:

From Figure~\ref{subfi:a} and Figure~\ref{subfi:b}, it can be observed that for both the classical shadow method and the direct quantum measurement method, there exists a clear boundary in the parameter space defined by the observable input methods LCP ($M, L, n, w$) and LHM ($M, k, n$), separating the regions where $t_{\text{shadow}} / t_{\text{direct}} \leq 1$ and $t_{\text{shadow}} / t_{\text{direct}} \geq 1$. This boundary precisely characterizes the resource consumption trade-offs between the two methods, revealing their relative advantages under different parameter regimes. Notably, for both observable input methods, it can be rigorously shown that $t_{\text{shadow}} / t_{\text{direct}}$ is independent of the value of $\epsilon$. The detailed proof is provided in the Appendix~\ref{sec:B}.
\\

\begin{table*}[htbp]  
\centering
\caption{Quantum Computer Performance: Measurement and Gate Times}
\label{tab:quantum_times}
\begin{tabular}{|l|c|c|c|c|}
\hline
\textbf{Quantum Computer Type} & \textbf{Measurement Time} & \textbf{Typical Range} & \textbf{Gate Time} & \textbf{Typical Range} \\ \hline
Superconducting (e.g., IBM)    & $10^{-5}\ \text{s}$                                      & 0.5--14~$\mu\text{s}$   & $10^{-8}\ \text{s}$                                    & 50--600 $\text{ns}$     \\ \hline
Ion Trap (e.g., IonQ)          & $10^{-4}\ \text{s}$                                      & 100--500 $\mu\text{s}$  & $10^{-5}\ \text{s}$                                    & 10--100 $\mu\text{s}$    \\ \hline
Photonic (e.g., Xanadu)        & $10^{-9}\ \text{s}$                                      & 1--10 $\text{ns}$       & $10^{-9}\ \text{s}$                                    & 0.1--10 $\text{ns}$      \\ \hline
Neutral Atom (e.g., Pasqal)    & $10^{-5}\ \text{s}$                                      & 10--50 $\mu\text{s}$    & $10^{-6}\ \text{s}$                                    & 0.5--5 $\mu\text{s}$     \\ \hline
\end{tabular}
\end{table*}

According to the intersection point in Figure~\ref{subfi:c}, when the observable takes the form of a linear combination of Pauli matrices, the classical shadow method begins to outperform the direct quantum measurement method at approximately \(\log(M) = 4\), corresponding to \(M \approx 16\), i.e., on the order of \(O(10^1)\). Due to parallel operations in quantum measurements, their efficiency is largely unaffected by either the number of qubits \(n\) or the Pauli weight \(w\). Although the number of measurements required by the classical shadow method scales logarithmically compared to that of the direct quantum measurement method, for small values of \(M\), the increased resource consumption caused by \(n\) and \(w\) outweighs the benefit gained from the reduced dependence on \(M\). This limits the ability of the classical shadow method on those scalings.

According to Figure~\ref{subfi:d}, which considers more general observables including all Hermitian matrices, we observe that when \(M\) is very small, as expected, the direct quantum measurement method outperforms the classical shadow method. Specifically, in the figure, when \(\log(M) \leq 6.5\), corresponding to \(M\) on the order of \(O(10^2)\), the direct quantum method exhibits a clear advantage. While one might anticipate that the classical shadow method would become dominant as \(M\) increases, an interesting reversal occurs around \(\log(M) = 26\), corresponding to \(M\) on the order of \(O(10^8)\), where the quantum measurement method again becomes more favorable. This is because, for observables represented as large matrices, the number of floating-point operations required by the classical approach grows exponentially with the number of qubits \(n\). 

It is also worth noting that, for observables in the form of linear combinations of Pauli matrices (LCP), the number of qubits \(n\) has only a minor impact on the resource consumption of the classical shadow method, indicating a weak dependence. In contrast, for observables represented as large Hermitian matrices, the resource consumption exhibits a stronger dependence on \(n\).

Figure~\ref{fig:big_figure} assumes a superconducting quantum computer as the default quantum device, with a quantum measurement time of approximately \(10^{-5}\, \text{s}\) and a quantum gate time of around \(10^{-8}\, \text{s}\). Additional data on quantum measurement and gate times for other types of quantum computers can be found in~\cite{AbuGhanem_2025}, including ion trap, photonic, and neutral atom quantum computers. The approximate measurement and gate times for these platforms, along with representative quantum computers for each type, are summarized in Table~\ref{tab:quantum_times}. We conducted analyses similar to those in Figure~\ref{subfi:c} and Figure~\ref{subfi:d} for the different types of quantum computers, plotting eight lines for each type of observable. These results are presented in Figure~\ref{subfig2:a} and Figure~\ref{subfig2:b}.

According to Figure~\ref{subfig2:a}, we observe that for LCP-type observables, the resource-saving advantage of the classical shadow method begins to emerge when \(M \approx 2^4 \sim 2^5\) (i.e., \(M = 16 \sim 32\)) for superconducting, ion trap, and neutral atom quantum computers. In contrast, for photonic quantum computers, this advantage becomes apparent only at \(M \approx 2^{6.5} \approx 100\), i.e., when \(M\) reaches \(\mathcal{O}(10^2)\). Overall, photonic quantum computers demonstrate significantly faster performance compared to other types of quantum computing platforms.

According to Figure~\ref{subfig2:b}, we observe that for LHM-type observables, the resource-saving advantage of the classical shadow method begins to manifest on superconducting, ion trap, and neutral atom quantum computers when \(M \approx 2^{6.5} \sim 2^{7.5}\), i.e., \(M \approx \mathcal{O}(10^2)\). As \(M\) increases further to approximately \(2^{26} \sim 2^{30}\) (i.e., \(M \approx \mathcal{O}(10^8 \sim 10^{10})\)), this advantage gradually diminishes, and the benefit of the direct quantum measurement method begins to emerge. In contrast, for photonic quantum computers, the performance curves of the classical shadow method and the quantum measurement method do not intersect, indicating that the implementation of the direct quantum measurement method on photonic platforms can be highly efficient.

\section{Methods}\label{sec3}
\hspace{1em} In this section, we present a method to estimate and compare the resources required for various types of observables, using both classical shadow techniques and  direct quantum measurement approaches. It is important to note that the primary objective of our research is to develop a cross-job resource estimation framework. Accordingly, in our numerical experiments, we employ ground state density matrices of randomly generated quantum states. This allows us to provide concrete numerical examples that validate the reliability and accuracy of our resource estimation method. In the following, we offer a concise overview of our approach to estimating the resource cost associated with the two types of observables.

\subsection{Resource Estimation for Observables in the Form of Linear Combinations of Pauli Matrices (LCP)}
\subsubsection{Classical Shadow}
\hspace{1em}In this scenario, resources are divided into classical and quantum components. We first randomly generate \( M \) observables for an \( n \)-qubit quantum system, where each observable is a linear combination of tensor products of Pauli matrices, consisting of \( L \) terms, each with a Pauli weight of \( w \). 
For example, for the observable 
$
O = \sum_{i=0}^{n-1} X_i X_{i+1} + Y_i Y_{i+1} + Z_i Z_{i+1}
$
, we have \( L = 3n \) and \( w = 2 \).

The quantum resource estimation process primarily involves rotating the initial quantum state using quantum gate operations and measuring the rotated state in the computational basis. During this rotation process, it is essential to determine the values of the single-qubit gates, which are directly related to the number of measurements required to achieve the desired precision. In our experiments, we assume a precision of $\varepsilon = 0.01$ and a failure probability of $\delta = 0.01$. The number of measurements required to estimate a single observable with Pauli weight $w$ is given in \cite{huang2020predicting}. However, we address a more general case where each observable must be decomposed into simpler Pauli observables of weight $w$. For each of these individual observables, the required number of measurements can be determined by following the procedure outlined in \cite{huang2020predicting}. By summing the measurement counts for all such decomposed observables, we obtain an estimate of the total number of measurements $T$ required for the general observable. Consequently, the total number of single-qubit gates is clearly $n \times T$, since each qubit must undergo a single-qubit gate operation during each measurement. According to \cite{AbuGhanem_2025}, for superconducting quantum computers, the operation time for a single-qubit gate is approximately $10^{-8}$ seconds, while a single quantum measurement takes around $10^{-5}$ seconds. Although Table 1 summarizes the corresponding data for various types of quantum computers, this study first takes superconducting quantum computers as an example to compare the two methods.

In classical computing, resource usage is typically quantified in terms of the number of floating-point operations (FLOPs). Classical computations involving matrix multiplications are central to many problems in the shadow method. However, when the observables are expressed in Pauli form, these calculations can be significantly simplified. Specifically, it suffices to verify whether the measurement basis matches the basis of the observable. If they match, the expectation value of the observable can be predicted by averaging the product of the corresponding eigenvalues, as detailed in the Appendix~\ref{sec:b1}. This approach eliminates the need for explicit matrix multiplications, thereby substantially reducing the classical computational resources required.
The rationale for distinguishing between different types of observables in this analysis is to optimize resource consumption based on the observable's structure. Accordingly, we estimate the number of FLOPs required for classical computations by analyzing the code used in this matching-based prediction method. Since our goal is to estimate resources rather than constrain the device to perform specific predictive operations, we can assume the maximum computational capability of the device.For this estimation, we assume the use of a supercomputer whose model and floating-point operation performance are listed on the TOP500 website \cite{top500_2024}. These systems typically operate at speeds on the order of $O(10^{14})$ to $O(10^{16})$ FLOPs per second. For our analysis, we assume a computational speed of approximately $10^{15}$ FLOPs per second.

\subsubsection{Direct Quantum Measurement}
An estimate of the number of quantum measurements required under specified conditions of precision and failure probability can be readily obtained using the Hoeffding inequality (a variant and corollary of the Chebyshev inequality). The left-hand side of the inequality represents the failure probability, $\delta$, at a given precision $\varepsilon$, while the right-hand side is a function of the number of measurements. Applying the Hoeffding inequality allows us to efficiently derive an upper bound on the required number of measurements.A key consideration is the distribution of error and failure probability among the terms in the observable. In the case of general Pauli-form observables, it is necessary to allocate the failure probability across each individual observable in the summation. The detailed derivation of the corresponding formula is provided in the Appendix~\ref{sec:b2}. Furthermore, Appendix~\ref{secD} presents a comparative example of two methods for measuring multi-body Pauli observables.

\subsection{Resource Estimation for Observables in the Form of Large Hermitian Matrices (LHM)}
\subsubsection{Classical Shadow}
The number of measurements required in this example differs due to the use of a different type of observable. To estimate the sufficient number of measurements, we must employ the shadow norm of the observable, as introduced in \cite{huang2020predicting}. However, this reference only provides a computational method for evaluating the shadow norm of Pauli-form observables. In our case, the shadow norm is replaced by the infinity norm, which can be shown to correspond to the worst-case scenario. In \cite{huang2020predicting}, an upper bound on the shadow norm is provided for observables expressed as linear combinations of Pauli operators, measured under Pauli measurement settings. However, this upper bound is not readily applicable to observables represented by large matrices unless these matrices are decomposed into Pauli form, which incurs significant additional computational cost.To address this, we demonstrate that when the sparsity $k \ll 2^n$, the upper bound can be directly approximated by the infinity norm. We then estimate this infinity norm using statistical methods, leading to an approximate upper bound expression that depends only on the sparsity $k$ and the number of qubits $n$. With this approximation, the computation of the shadow norm becomes straightforward. By substituting the estimated upper bound into the expression for the required number of measurements $N$, we can effectively estimate the quantum resources needed for the task.

The classical computational resources required are comparable to those for Pauli-form observables. However, when dealing with large Hermitian matrix observables, matrix multiplications become unavoidable. To reduce resource consumption and accelerate computations, we employ sparse matrix techniques. As in previous cases, the number of floating-point operations (FLOPs) is determined by implementing and analyzing the corresponding computational procedures in code.

\subsubsection{Direct Quantum Measurement}
In this case, we encounter the problem of estimating norms related to the observable. When measuring a large matrix observable \( O \), each measurement returns one of its eigenvalues, which lies within \( [-\|O\|_2, \|O\|_2] \), where \( \|O\|_2 \) is the spectral norm (the largest absolute eigenvalue). According to the Hoeffding inequality:
\begin{align}
P(|\bar{X} - \langle O \rangle| \geq \varepsilon) \leq 2 \exp \left( -\frac{2N \varepsilon^2}{(b - a)^2} \right)
\end{align}
Here, $N$ denotes the number of measurements for a single observable, and the right-hand side contains $b - a = 2 \|O\|_2$, where $\|O\|_2$ is the spectral norm of the Hermitian matrix $O$. To proceed, the spectral norm must be estimated in terms of the matrix's sparsity $k$. The square of the spectral norm admits a natural upper bound related to both the 1-norm and the infinity norm of the matrix. For Hermitian matrices, the 1-norm and infinity norm are equal, so we cannot obtain a tighter upper bound on the infinity norm beyond what can be derived from the statistical distribution of the matrix elements. Finally, by substituting this bound into the Hoeffding inequality, we obtain the required number of measurements.

\hspace{1em}Thus, we conclude the estimation of computational resources required for both types of observables under different methods. The detailed calculations and derivations are provided in the Appendix~\ref{sec:B}.

In physical experiments, observables in the LHM form can often be measured by first decomposing them into the LCP form. This specific decomposition method is detailed in Appendix~\ref{secC}. While this approach is frequently feasible, the resource cost associated with such decomposition can grow exponentially in many other scenarios. Consequently, alternative measurement strategies are considered in experiments. Common methods include mapping the expectation value onto a simple observable for direct readout via ancillary-qubit interference, such as the Hadamard test or the SWAP test, or through interference and collective measurement on two copies. A typical example is the direct measurement of the second-order {\color{black}R\'enyi} entropy using the SWAP test~\cite{PhysRevLett.104.157201,Brydges_2019}. We provide a detailed resource estimation for this experiment in Appendix~\ref{secD}.

\section{Conclusion}\label{sec4}
\hspace{1em}In this paper, we have conducted a comprehensive resource estimation of the classical shadow technique, encompassing both classical and quantum components. Unlike existing approaches, which often rely on empirical or simulation-based analysis, our study provides an analytical and quantitative framework for resource estimation, which can more accurately predict the consumed resources.

Our results reaffirm that the classical shadow method achieves logarithmic savings in the number of measurements. However, it does not show a significant advantage over direct quantum measurement when observables are either extremely simple or highly complex and irregular. This clarifies the method’s practical scope. Specifically, under typical assumptions, the method is advantageous when: (i) the number of observables expressed as linear combinations of Pauli operators is large, or (ii) the number of large Hermitian-matrix observables falls within a certain range. As our estimates are based on worst-case bounds, practical resource costs are likely lower.

This work offers practical guidance for deploying quantum measurement techniques. A key remaining challenge is the development of an adaptive switching mechanism between classical shadow and direct measurement, especially in dynamic scenarios such as real-time quantum system monitoring. Furthermore, classical simulation of both methods allows for pre-optimized measurement strategies, thereby reducing experimental overhead. Additional contributions include a detailed comparative analysis of physical property extraction, theoretical upper bounds under realistic constraints, {\color{black}and a discussion on observable formulation, all of which enhance the generality of the approach;} see the Appendix for details.
{\color{black}
In this work, our comparisons and analysis are mainly confined to the NISQ framework. 
Under the fully fault-tolerant quantum computing (FTQC) paradigm, however, quantum error 
correction introduces additional resource overheads that are not captured by the NISQ-level 
runtime model. For example, in surface-code-based architectures~\cite{2024}, encoding one 
logical qubit generally requires many physical qubits in order to suppress the logical error 
rate to an algorithmically relevant level. Moreover, a logical Pauli measurement is typically 
implemented through repeated syndrome extraction over multiple error-correction cycles, so 
its runtime can be substantially longer than that of a corresponding physical measurement.

To give an order-of-magnitude comparison, one may introduce an effective FTQC overhead factor 
\(C_{\mathrm{FT}}\). This factor should be understood as an architecture-dependent effective 
quantity rather than a universal constant. In particular, for surface-code-based implementations, 
the effective overhead may depend on the chosen code distance, the logical operation schedule, 
the throughput of magic-state distillation when non-Clifford operations are required, decoder 
latency, and the target logical failure probability. These quantities may also vary with the 
circuit size, circuit depth, and the structure of the observables being measured.

Within this coarse-grained interpretation, \(C_{\mathrm{FT}}\) provides a simplified way to 
indicate how fault-tolerant implementation may enlarge the per-round runtime. Classical shadows 
rely on implementing random Pauli-basis transformations over logical qubits, and the resulting 
logical circuits may be less efficient than optimized direct-measurement schemes tailored to 
specific observables, such as schemes based on Pauli-term grouping. The latter measurement 
circuits are typically deterministic and may have smaller logical depth. Consequently, although 
classical shadows can reduce the number of measurements required, the increased logical 
measurement time in an FTQC implementation may partially offset this advantage and may shift the 
system size or precision regime in which classical shadows outperform direct quantum measurement 
toward larger scales.

Therefore, the FTQC-related discussion in this work should be regarded as a qualitative 
order-of-magnitude estimate rather than a complete architecture-level resource analysis. A more 
precise FTQC runtime comparison would require specifying a concrete error-correction architecture, 
logical compilation strategy, decoder model, and target logical failure probability.
}

In our runtime analysis, we consider the time costs of measurement, gate operations, and classical post-processing. The focus of this work is to compare the efficiency of two methods in the information-extraction stage after a given quantum state has been prepared. This separated analysis aligns with the experimental paradigm in the NISQ era, where state preparation and measurement are often independently optimized modules. Therefore, we do not include the resource overhead of quantum state preparation in our main comparison, but we provide a brief analysis of the overall resource consumption when state preparation is incorporated.

We model the preparation cost as a constant term \( T_{\text{prep}} \) and include it in the total runtime framework. Specifically, for a protocol requiring \( N \) rounds of measurements, the total runtime can be expressed as
\[
T_{\text{total}} = N \cdot \left( T_{\text{prep}} + T_{\text{meas}} + T_{\text{gate}} \right) + T_{\text{classical}},
\]
where \( T_{\text{prep}} \) denotes the time required for a single round of quantum state preparation, {\color{black}\( T_{\text{meas}} \) denotes the time required to perform one measurement round, \( T_{\text{gate}} \) denotes the time required to implement the quantum gate operations involved in one round of the protocol, and \( T_{\text{classical}} \) denotes the classical post-processing time after collecting the measurement outcomes. Here, \(N\) denotes the total number of repeated measurement rounds required by the corresponding protocol.}

Importantly, when each measurement round requires preparing the same target state, \( T_{\text{prep}} \) appears as a shared additive cost per repetition for both classical shadows and direct quantum measurement. In this case, the dominant preparation contribution \( N \cdot T_{\text{prep}} \) acts as a common runtime contribution for both methods. Although it increases the absolute runtime, it does not modify the relative comparison between the two strategies, nor does it affect the crossover behavior identified in our analysis. Therefore, as long as the preparation cost per round remains method-independent, including it does not change the qualitative scaling conclusions of this work.

This modeling is further supported by practical considerations. In many quantum systems, specific state preparations can be realized via highly optimized circuits with effectively constant time overhead. For example, in photonic quantum computing systems, Bell-state preparation can be accomplished with linear optical elements in constant time (typical timescales on the order of \(10\)--\(100\) nanoseconds) \cite{Natarajan_2012,thiele2024cryogenicfeedforwardphotonicquantum,Maeder_2026}. Such preparation processes do not scale with system size and can be executed in parallel across multiple qubits.

We note, however, that for certain complex quantum states (e.g., ground states prepared via variational or adiabatic procedures), the preparation cost may grow with the system size. In such cases, state preparation may dominate the overall runtime. Nevertheless, provided that both measurement strategies require the same number of independent state re-preparations, the preparation cost contributes equally to both methods and does not qualitatively alter their comparative analysis. Incorporating detailed scaling models for state preparation into a fully unified resource framework remains an interesting direction for future work.

In future work, we plan to extend the present analysis beyond Pauli measurements by incorporating Clifford measurement settings, which would make the resource comparison between classical shadows and direct quantum measurement more comprehensive. In addition, comparing the energy consumption of the two methods within the same resource-estimation framework would make the analysis more relevant for energy-efficient quantum computing~\cite{tu2025towards}. 

\section*{Acknowledgment}

We thank Hsin-Yuan Huang for helpful discussions.

\vspace{7cm}

\clearpage
\bibliography{ref}
\pagebreak

\onecolumn
\appendix

\begin{center}
	{\Large \bf Appendix}
\end{center}

\section{Introduction to Classical Shadow method }\label{secA}

\subsection{Construction of Classical Shadows}
The classical shadow described in the paper \cite{huang2020predicting} is a classical description of a quantum state, which is created by special measurements and further processing quantum state. Data acquisition and processing are the challenges with two major effort-intensive steps of the construction process.\\
\\
In a multi-qubit system, for an unknown quantum state $\rho$, randomly select a unitary transformation $U$ from a fixed unitary set $\mathcal{U}$, and rotate the quantum state as $\rho\mapsto U\rho U^{\dagger}$. Subsequently, perform a measurement on the rotated quantum state in the computational basis $\{|b\rangle: b\in\{0,1\}^n\}$. According to Born's rule, the probability of obtaining measurement result $|\hat{b}\rangle\in\{0,1\}^n$ is $\text{Pr}[\hat{b} = b]=\langle b|U\rho U^{\dagger}|b\rangle$. After measurement, process the result and store $U^{\dagger}|\hat{b}\rangle\langle\hat{b}|U$. This random measurement outcome $U^{\dagger}|\hat{b}\rangle\langle\hat{b}|U$ contains valuable information about $\rho$ in expectation, with $\mathbb{E}[U^{\dagger}|\hat{b}\rangle\langle\hat{b}|U]=\mathcal{M}(\rho)$, where $\mathcal{M}$ is a quantum channel determined by the unitary set $\mathcal{U}$.\\
\\
We can now construct the classical shadow. To extract quantum state information from measurements, further process $U^{\dagger}|\hat{b}\rangle\langle\hat{b}|U$. Since $\mathcal{M}$ as a linear map has a unique inverse (provided the measurement defined by $\mathcal{U}$ is tomographically complete), denoted as $\mathcal{M}^{-1}$. Applying $\mathcal{M}^{-1}$ to $U^{\dagger}|\hat{b}\rangle\langle\hat{b}|U$ yields $\hat{\rho}=\mathcal{M}^{-1}(U^{\dagger}|\hat{b}\rangle\langle\hat{b}|U)$, which is a classical shadow of $\rho$. Repeating this process $N$ times generates $N$ independent classical shadows, forming an array:
\begin{align}
S(\rho; N)=\{\hat{\rho}_1=\mathcal{M}^{-1}(U_1^{\dagger}|\hat{b}_1\rangle\langle\hat{b}_1|U_1),\ldots,\hat{\rho}_N=\mathcal{M}^{-1}(U_N^{\dagger}|\hat{b}_N\rangle\langle\hat{b}_N|U_N)\}
\end{align}
This completes the classical shadow construction.

\subsection{Algorithm capable of predicting expectations of observables}
Algorithm 1 from \cite{huang2020predicting} is a median-of-means prediction algorithm based on classical shadows $S(\rho, N)$, designed to predict linear function values of quantum states. The specific process is:\\
\\
The function takes three parameters: observables $O_1,\ldots,O_M$, classical shadows $S(\rho; N) = [\hat{\rho}_1,\ldots,\hat{\rho}_N]$, and a positive integer $K$. Here, $S(\rho; N)$ contains $N$ classical approximations of $\rho$, and $K$ determines how to partition the shadows for robust estimation.\\
\\
The first step is dividing $S(\rho; N)$ into $K$ equal parts. For each part, compute sample means to build $K$ estimators:
\begin{align}
\hat{\rho}^{(k)} = \frac{1}{\lfloor N/K \rfloor} \sum_{i=(k - 1)\lfloor N/K \rfloor + 1}^{k\lfloor N/K \rfloor} \hat{\rho}_i, \quad k = 1,\ldots,K
\end{align}
This partitioning mitigates outlier effects by creating multiple independent estimators.
\\
\\
The second step is using the median-of-means Algorithm. For each observable $O_i\ (i = 1,\ldots,M)$, compute traces $\text{tr}(O_i\hat{\rho}^{(k)})$ across all $K$ estimators. The final prediction is the median of these values:
\begin{align}
\hat{o}_i(N, K) = \text{median}\{\text{tr}(O_i\hat{\rho}^{(1)}),\ldots,\text{tr}(O_i\hat{\rho}^{(K)})\}
\end{align}
Median estimation enhances robustness against outliers compared to simple averaging.\\
\\
Through these steps, Algorithm 1 effectively predicts linear function values of quantum states using classical shadows and median-of-means techniques.

\section{Scaling analysis}\label{sec:B}
\subsection{Classical shadow method (LCP): proof of Statement 1}
\label{sec:b1}
Restate the definitions of some symbolically represented parameters:
\begin{itemize}
    \item $M = \texttt{num\_observables}$
    \item $L = \texttt{terms\_per\_observable}$
    \item $w = \texttt{pauli\_weight}$
    \item $T = \texttt{max\_measurements}$
    \item $K = \texttt{num\_groups}$
    \item $\varepsilon = \texttt{estimation\_error}$
    \item $\delta = \texttt{failure\_probability}$
    \item $k = \texttt{sparsity\_of\_matrix}$
\end{itemize}
\subsubsection{The estimation of quantum resources}
In this case, the formula for the total number of required measurements along with its proof, as well as the calculation method for $ \| O \|_{\text{shadow}}^2$ under these conditions, can be found in the appendix of \cite{huang2020predicting} and are listed as follows:
\begin{align}
 T = \left\lceil \frac{34}{\varepsilon^2} \cdot \max_i \| O_i \|_{\text{shadow}}^2 \cdot 2 \cdot \log \left( \frac{2M}{\delta} \right) \right\rceil 
 \end{align}
We can find $\| O_i \|_{\text{shadow}}^2=3^w$ in \cite{huang2020predicting} when L equals 1 and there is no coefficient preceding each observable. For the general case, $\| O_i \|_{\text{shadow}}^2 = \sum_{j = 1}^L c_j^2 \cdot 3^w$. Moreover, in Statement 1 we assumed $c_j \sim \mathcal{N}(0, 0.5)$ truncated to $[-1, 1]$, so we can easily calculate the average $c_j^2 \approx 0.25$. Therefore, $\| O_i \|_{\text{shadow}}^2 \approx L \cdot 0.25 \cdot 3^w$.\\
\\
Substituting the aforementioned approximation of $\| O_i \|_{\text{shadow}}^2$, we obtain the following expression:
\begin{align}
T &\lesssim \frac{17L \cdot 3^{w}}{\varepsilon^{2}} \cdot \log\left(\frac{2M}{\delta}\right)
\end{align}
The quantum resource component in Statement 1 has been rigorously proven.
\subsubsection{The estimation of classical resources}
For resource estimation of classical computations, we use the number of floating-point operations as a proxy, and only need to analyze the floating-point operations in the code.\\
\\
\textbf{Core Function Calls}\\
The program uses the Classical Shadow method to estimate the expectation values of complex observables. Some Core functions related to classical computing are as follows::
\begin{itemize}
    \item \texttt{randomized\_classical\_shadow}: Generates random measurement bases
    \item \texttt{estimate\_exp}: Processes a single Pauli term
    \item \texttt{estimate\_exp\_mom}: Performs grouped estimation
    \item \texttt{estimate\_complex\_observable}: Processes all Pauli terms for a single observable
    \item \texttt{estimate\_multiple\_observables}: Estimates all observables
\end{itemize}
In the following, we will divide the method of using the classical shadow approach to predict the expectation of observables in the form of linear combinations of Pauli operators into six steps. Step 1 first randomly generates the measurement bases; Step 2 simulates quantum measurements; Step 3 estimates the expectation of a single term; Step 4 applies the Median-of-means algorithm on the basis of Step 3; Step 5 sums each individual term to obtain the expectation of each observable; Step 6 predicts the expectations of all observables. The functions used in each step are presented in the form of Python code, along with a brief explanation and an analysis of floating-point operations.\\
\\
\vspace{0.3cm}
\textbf{Step 1: Generation of Measurement Basis}\\
\begin{figure*}[htbp]
    \centering 
    \includegraphics[width=1.15\textwidth, trim=60pt 700pt 10pt 60pt, clip]{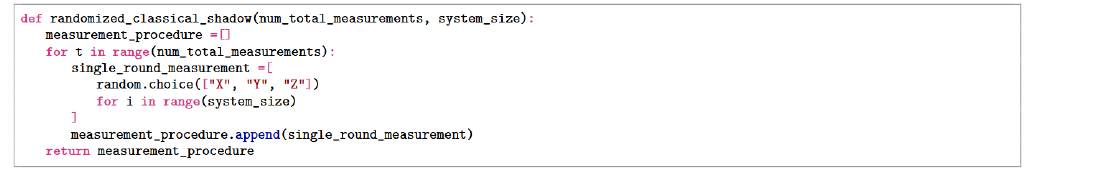}
    \captionsetup{margin={0cm,0cm}} 
    \caption{\textbf{Function: \texttt{randomized\_classical\_shadow}}}  
    \label{fig:code1} 
\end{figure*}
In Step 1, we construct a function for randomly generating measurement bases, which assigns X/Y/Z bases to different qubits with uniform probability. As we can see in Figure~\ref{fig:code1}, regardless of the size of the quantum system, the measurement process is performed independently for each individual qubit. Since the number of required measurements is $T$, the number of measurement bases to be generated is $nT$. However, this step does not involve floating-point operations and thus is not counted. Next, we calculate the matching probability. Briefly speaking, it refers to the probability that the measurement basis at this position is the same as the Pauli operator at the corresponding position of the observable. This matching probability plays an important role in Step 3.\\
\\
First, we consider the calculation of matching probability, for a Pauli-form observable with only one term and a Pauli weight of $w$(e.g. $Z_0X_2Y_5$, with $w=3$), the single-position matching probability: $P(\text{basis} = \text{operator}) = \frac{1}{3}$. Therefore, the probability that all $w$ positions match simultaneously:
\begin{align}
     p = \left( \frac{1}{3} \right)^w
\end{align}
\textbf{Step 2: Quantum Measurement}\\
In this step, we use a classical computer to simulate quantum computer measurements and store the results of $T$ measurements in $\text{full\_measurement}$. This step does not involve any classical computing and naturally does not involve any floating-point operations.\\
\\
\textbf{Step 3: Estimate of Expectation}\\
\begin{figure*}[htbp]
    \centering 
    \includegraphics[width=1.2\textwidth, trim=70pt 680pt 10pt 40pt, clip]{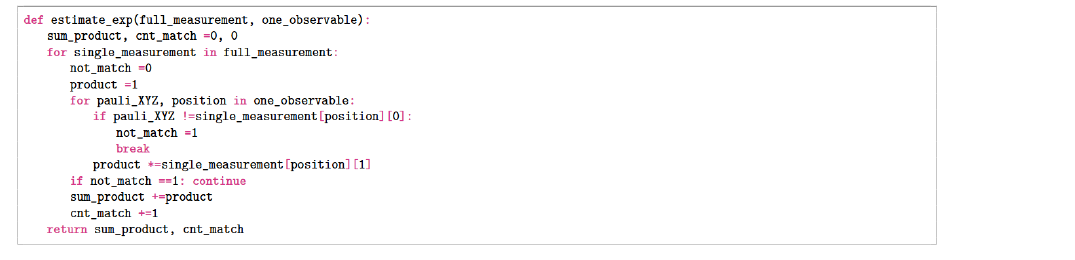}
    \captionsetup{margin={0cm,0cm}} 
    \caption{\textbf{Function: \texttt{estimate\_exp}}}  
    \label{fig:code2} 
\end{figure*}
\vspace{-0.3cm} 
In Step 3, we use a function from the code attached in the original paper that proposed classical shadow \cite{huang2020predicting}. This code is very concise and efficient, avoiding complex matrix multiplications in classical shadows and directly using a matching method to predict the expectation. The term "one\_observable" in the function represents one of the terms in the summation of an observable expressed as a linear combination of Pauli operators, and the summation will be performed in Step 5. The specific principle of this code will not be explained in detail; we only briefly describe the calculation of floating-point numbers:
\begin{itemize}
    \item \textbf{Input}: $T'$ measurements (group size), $w$ Pauli operators
    \item \textbf{Core operations}: This function includes three major steps.
Checking whether basis matches is the first step, which are all boolean operations. Boolean operations not counted as FLOPs. The sencond and third steps are Floating-point multiplication only when bases match and Floating-point addition only when all positions match. These two steps include some operations involving floating-point arithmetic.

    \item \textbf{Total floating-point operations:} 
The complete match probability is \( p = (1/3)^w \). In such a scenario, the floating-point operations involved consist of \( w \) multiplications (when a match occurs) plus 1 addition (specific to the complete match). Based on this composition, the expected number of floating-point operations can be expressed as
\begin{align}
    \text{FLOPs}_{\text{exp}} = T' \cdot p \cdot (w + 1) 
\end{align}
where the term \( (w + 1) \) directly corresponds to the total count of these multiplicative and additive operations.
\end{itemize}
\vspace{0.3cm} 
\textbf{Step 4: Using the Median of Means Method}

\begin{figure*}[htbp]
    \centering 
    \includegraphics[width=1.2\textwidth, trim=75pt 690pt 5pt 50pt, clip]{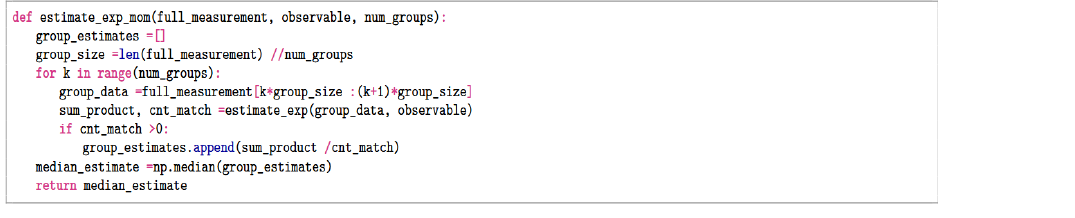}
    \captionsetup{margin={0cm,0cm}} 
    \caption{\textbf{Function: \texttt{estimate\_exp\_mom}}}  
    \label{fig:code3} 
\end{figure*}

In Step 4, we use the Median-of-means algorithm to enhance the accuracy and stability of the results. In practice, it adds operations on the basis of \textit{estimate\_exp}, specifically grouping first, then taking the average of each group, and finally taking the median. The following is the floating-point number analysis for this function:
\begin{itemize}
\item \textbf{Input}: $T$ measurements, $K$ groups
\item \textbf{Core operations}:
The process involves calling the \texttt{estimate\_exp} function $K$ times, with each call utilizing $T' = T/K$ measurements. That is, the total number of measurements $T$ is divided into $K$ groups, each of size $T'$. Within each group, a key operation is a division corresponding to one floating-point operation, which is executed when a match condition is met. Following these group-level computations, the median calculation is performed to aggregate the results across groups. This step mainly consists of comparison operations, and the associated floating-point cost is negligible in the overall complexity assessment.
    \item \textbf{Total floating-point operations}:
  \begin{align}
    \text{FLOPs}_{\text{mom}} = K \cdot \left[ \frac{T}{K} \cdot p \cdot (w + 1) + \mathbb{I}_{\text{match}} \right] = T \cdot p \cdot (w + 1) + K_{\text{eff}}
  \end{align}
$K_{\text{eff}}$ is the number of groups with actual matches ($\leq K$), conservatively estimated as $K$
\end{itemize}
\vspace{0.3cm} 
\textbf{Step 5: Summing the Expectation of Each Term}

\begin{figure*}[htbp]
    \centering 
    \includegraphics[width=1.15\textwidth, trim=60pt 710pt 5pt 55pt, clip]{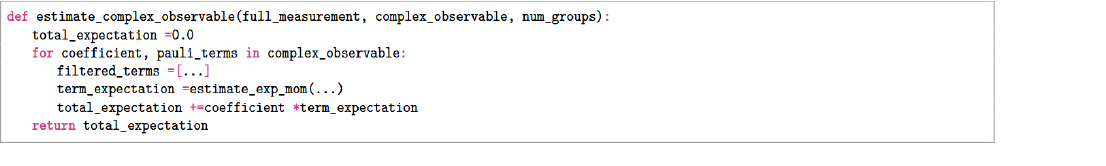}
    \captionsetup{margin={0cm,0cm}} 
    \caption{\textbf{Function: \texttt{estimate\_complex\_observable}}}  
    \label{fig:code4} 
\end{figure*}
\vspace{-0.3cm} 

Since Step 3 and Step 4 only estimate the expectation of a single-term Pauli-form observable, we sum the expectations of each Pauli-form observable term in Step 5 to obtain the total expectation. The following is a brief analysis of floating-point operations:
\begin{itemize}
    \item \textbf{Input}: $L$ Pauli terms
    \item \textbf{Core operations}:
The floating-point estimation of this function is relatively simple; it merely involves calling the previous \textit{estimate\_exp\_mom} function L times. During each call, two additional floating-point operations are required: the first is a multiplication needed to construct the weight (coefficient) preceding each Pauli term, and the second is the summation of each individual term.
    \item \textbf{Total floating-point operations}:
    \begin{align}
    \text{FLOPs}_{\text{complex}} = L \cdot \left[ T \cdot p \cdot (w + 1) + K + 2 \right]
    \end{align}
\end{itemize}
\vspace{0.3cm} 
\textbf{Step 6: Predict the Expectations of All Observables}

\begin{figure*}[htbp]
    \centering 
    \includegraphics[width=1.15\textwidth, trim=60pt 710pt 5pt 65pt, clip]{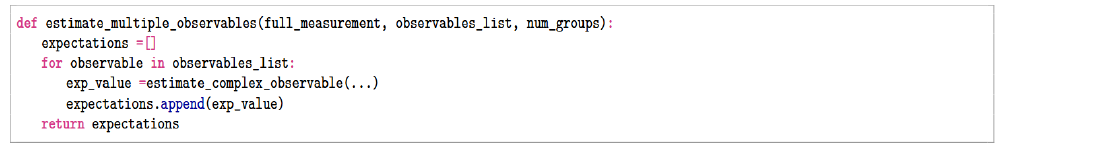}
    \captionsetup{margin={0cm,0cm}} 
    \caption{\textbf{Function: \texttt{estimate\_multiple\_observables}}}  
    \label{fig:code5} 
\end{figure*}

In this step, we will include all observables and need to introduce the number of observables $M$. We apply the previous Steps 3, 4, and 5 to each observable to predict the expectation. The analysis of floating-point operations is simply a multiplication by $M$, as follows:
\begin{itemize}
    \item \textbf{Input}: List of $M$ observables
    \item \textbf{Core operations}: Calls \texttt{estimate\_complex\_observable} $M$ times
    \item \textbf{Total floating-point operations}:
   \begin{align}
    \text{FLOPs}_{\text{total}} = M \cdot L \cdot \left[ T \cdot \left( \frac{1}{3} \right)^w \cdot (w + 1) + K + 2 \right]
   \end{align}
\end{itemize}
Finally, substituting the value of $K$ yields the final result:
\begin{align}
\text{FLOPs}_{\text{total}} &\lesssim M \cdot L \cdot \left( T \cdot \left( \frac{1}{3} \right)^w \cdot (w + 1) + 2 \cdot \log\left( \frac{2M}{\delta} \right) + 2 \right) 
\end{align}
With this, the proof of Statement 1 is complete.

\subsection{Direct Quantum Measurement} (LCP): proof of Statement 2
\label{sec:b2}
In quantum mechanics, directly measuring the expectation value \(\text{Tr}(O_m\rho)\) of an observable \(O_m\) (where \(\rho\) is the quantum state, \(m = 1, 2, \ldots, M\)) typically involves projecting the quantum state onto the eigenbasis of \(O_m\), performing multiple measurements, and averaging the results.  For observables expressed as linear combinations of Pauli operators, \( O = \sum_{j=1}^L c_j P_j \), this method measures each Pauli term \( P_j \) individually and combines the results to compute \( \langle O \rangle \).The resource estimation of this method is achieved through the following five steps:\\
\\
\textbf{Step 1: Approximate the Maximum Sum of Absolute Coefficients}
\begin{align}
\text{max\_sum\_abs\_coeff} = L \cdot \mu
\end{align}
Each observable has \( L \) Pauli terms with coefficients \( c_j \). The sum \( \sum_{j=1}^L |c_j| \) scales the error in the expectation value. Here, \( \mu \) (default 0.5) approximates the average \( |c_j| \), so \( L \cdot \mu \) estimates this sum.\\
\\
\textbf{Step 2: Allocate Error and Failure Probability}
\begin{align}
\epsilon' = \frac{\epsilon}{\text{max\_sum\_abs\_coeff}} = \frac{\epsilon}{L \cdot \mu}
\end{align}
\begin{align}
\delta' = \frac{\delta}{M \cdot L}
\end{align}
\begin{itemize}

    \item \textbf{Error (\( \epsilon' \))}: The total error \( \epsilon \) is divided by \( L \cdot \mu \) to allocate an error per Pauli term.
    \item \textbf{Failure Probability (\( \delta' \))}: The total \( \delta \) is split across \( M \cdot L \) terms using the union bound.
\end{itemize}
\textbf{Step 3: Measurements per Pauli Term}
\begin{align}
N_j = \left\lceil \frac{2}{\epsilon'^2} \ln\left(\frac{2}{\delta'}\right) \right\rceil
\end{align}
Using the Hoeffding inequality, for Pauli outcomes \( \pm 1 \), the number of measurements \( N_j \) ensures the error is within \( \epsilon' \) with probability \( 1 - \delta' \):
\begin{align}
\delta'=P(|\bar{X} - \langle P_j \rangle| \geq \epsilon') \leq 2 \exp\left(-\frac{N_j \epsilon'^2}{2}\right) 
\end{align}
\begin{align}
N_j \leq  \frac{2}{\epsilon'^2} \log\left(\frac{2}{\delta'}\right)
\end{align}
\\
\textbf{Step 4: Total Number of Measurements}
\begin{align}
T' = M \cdot L \cdot N_j
\end{align}
With \( M \) observables and \( L \) terms each, the total measurements are the product.\\
\\
\textbf{Step 5: Final Expression}
\begin{align}
T' \leq  M \cdot L \cdot \left\lceil \frac{2}{\left( \frac{\epsilon}{L \cdot \mu} \right)^2} \log\left( \frac{2}{\frac{\delta}{M \cdot L}} \right) \right\rceil 
\end{align}
Simplified:
\begin{align}
T' \leq  M \cdot L \cdot \left\lceil \frac{2 (L \cdot \mu)^2}{\epsilon^2} \log\left( \frac{2 M \cdot L}{\delta} \right) \right\rceil 
\end{align}
Substituting the default value $\mu=0.5$ yields:
\begin{align}
T' &\lesssim  \frac{0.5ML^3 }{\epsilon^2} \log\left( \frac{2ML}{\delta} \right)
\end{align}
With this, the proof of Statement 2 is complete.\\
\\
Before proving Statements 3 and 4, we first present some important remarks.

\subsection{Estimation of shadow norm and quantum resources for Large Matrices method: proof of Statement 3(a)}
\label{sec:b3}
Since the shadow norm is a crucial norm in the classical shadow method, its magnitude directly determines the number of measurements required.For observables in the form of large matrices, we can use their infinity norm instead of the shadow norm. The following are explanations and justifications for this approach:\\
\\
In \cite{huang2020predicting}, the shadow norm is defined as
\begin{align}
    \| O \|_{\text{shadow}}^2 = \max_{\sigma \text{ (state)}} \mathbb{E}_{U \sim \mathcal{U}} \sum_{b \in \{0,1\}^n} \langle b | U \sigma U^\dagger | b \rangle \cdot [\text{Tr}(O \mathcal{M}^{-1}(U^\dagger | b \rangle \langle b | U))]^2
\end{align}
The shadow norm in numerical computation for drawing images is defined as follows:
\begin{align}
 \left\| O - \frac{\text{Tr}(O)}{d} I \right\|_{\text{shadow}}^2 = \left\| O - \frac{\text{Tr}(O)}{d} I \right\|_{\infty}^2
\end{align}
where:
\begin{itemize}
    \item $d = 2^n$: the dimension of the Hilbert space ($n$ is the number of qubits),
    \item $\text{tr}(O)$: the trace of the observable,
    \item $\| \cdot \|_{\infty}$: the infinity norm, i.e., the largest singular value of the matrix.
\end{itemize}
\textbf{Simple verbal explanations of this claim}:\\
First, state an important conclusion:  
For the random Pauli measurement scheme, the theoretical derivation in \cite{huang2020predicting}reveals a key result: The upper bound of the shadow norm $\| O \|_{\text{shadow}}^2$ is directly related to the infinity norm of the observable $O$, specifically expressed as:
\begin{align}
\| O - \frac{\text{tr}(O)}{d} I\|_{\text{shadow}}^2 \leq c \cdot \left\| O - \frac{\text{tr}(O)}{d} I \right\|_{\infty}^2
\end{align}

\begin{itemize}
     \item where $c$ is a constant related to the measurement scheme and the form of the observable (for random Pauli measurements, when the observable is a linear combination of Pauli matrices, $c=4^w$, where $w$ is the Pauli weight).

    \item When \( w \ll n \), i.e., \( 2^w \ll 2^n \), we consider directly using the infinity norm to replace the shadow norm. (Since the sparsity \( k \) selected in our large matrix operations is very small compared to \( 2^n \), the matrices we study are relatively local. Alternatively, in the Pauli linear combination form obtained by decomposing a large matrix, the Pauli weight \( w \) is much smaller than \( n \). Moreover, it is shown in \cite{huang2020predicting} that the upper bound \( 4^k \| O \|_{\infty} \) is a relatively sparse upper bound. Therefore, in numerical calculations, the computational complexity mainly arises from the observable \( O \) itself. Hence, it is reasonable to ignore constant terms.)
\end{itemize}
\subsubsection{Estimation of \(\left\|O - \frac{\text{tr}(O)}{2^n}I\right\|_{\infty}\)}
We need to estimate the value of \(\left\|O - \frac{\text{tr}(O)}{2^n}I\right\|_{\infty}\), where \(O\) is a \(2^n\times2^n\) Hermitian matrix with each row containing \(k\) non-zero elements, and these non-zero elements follow a standard normal distribution \(N(0, 1)\). Here, \(n\) is the number of qubits, \(I\) is the identity matrix, and \(\text{tr}(O)\) is the trace of \(O\). We aim to obtain an expression in terms of \(k\) and \(n\) to roughly describe the magnitude of this norm.\\
\\
First, recall the definition of \(\|A\|_{\infty}\). For a matrix \(A\), the infinity norm is the maximum of the sums of absolute values of the rows:
\begin{align}    
\|A\|_{\infty} = \max_{1\leq i\leq 2^n} \sum_{j = 1}^{2^n} |a_{ij}|
\end{align}
Therefore, our goal is to compute:
\begin{align}  
\left\|O - \frac{\text{tr}(O)}{2^n}I\right\|_{\infty} = \max_{1\leq i\leq 2^n} \sum_{j = 1}^{2^n} \left|o_{ij} - \frac{\text{tr}(O)}{2^n}\delta_{ij}\right|
\end{align}  
where \(\delta_{ij} = 1\) if \(i = j\), and 0 otherwise. This implies that for each row \(i\):
\begin{align} 
\sum_{j = 1}^{2^n} \left|o_{ij} - \frac{\text{tr}(O)}{2^n}\delta_{ij}\right| = \sum_{j\neq i} |o_{ij}| + \left|o_{ii} - \frac{\text{tr}(O)}{2^n}\right|
\end{align} 
We need to estimate the maximum of this sum.\\
\\
Then, we list the properties of $O$ below.
\begin{enumerate}
    \item \textbf{Hermiticity}: \(O = O^{\dagger}\). Since \(O\) is a real matrix (elements from \(N(0, 1)\)), this means \(O\) is symmetric, i.e., \(o_{ij} = o_{ji}\).
    \item \textbf{Sparsity}: Each row has \(k\) non-zero elements, and due to symmetry, each column also has \(k\) non-zero elements.
    \item \textbf{Element Distribution}: Non-zero elements \(o_{ij}\) are drawn from the standard normal distribution \(N(0, 1)\).
\end{enumerate}
One of our goals is to estimate $\text{tr}(O)$. The trace is the sum of all diagonal elements:
\begin{align} 
\text{tr}(O) = \sum_{i = 1}^{2^n} o_{ii}
\end{align} 
However, \(o_{ii}\) are not necessarily non-zero. Each row has \(k\) non-zero elements randomly selected (without replacement) from \(2^n\) positions, so the probability that \(o_{ii} \neq 0\) is \(\frac{k}{2^n}\).
\begin{itemize}
    \item If \(o_{ii} \neq 0\), then \(o_{ii} \sim N(0, 1)\);
    \item If \(o_{ii} = 0\), it does not contribute to the trace.
\end{itemize}
The number of non-zero diagonal elements approximately follows a binomial distribution \(\text{Bin}(2^n, \frac{k}{2^n})\), with expectation \(k\). Each non-zero \(o_{ii} \sim N(0, 1)\), so:
\begin{align} 
\text{tr}(O) = \sum_{i: o_{ii}\neq 0} o_{ii}
\end{align} 
Its variance is:
\begin{align} 
\text{Var}(\text{tr}(O)) = \text{expected number of non-zero terms} \times \text{Var}(o_{ii}) = k \cdot 1 = k
\end{align} 
Thus, \(\text{tr}(O) \sim N(0, k)\), and:
\begin{align} 
\frac{\text{tr}(O)}{2^n} \sim N\left(0, \frac{k}{2^{2n}}\right)
\end{align} 
For large \(n\), \(\frac{\text{tr}(O)}{2^n}\) is very small, with a standard deviation of \(\sqrt{\frac{k}{2^{2n}}}\). For example, if \(k\) is constant or grows slowly with \(n\) (e.g., \(k = n\)), this term can be neglected compared to others.
\vspace{0.3cm}
\textbf{Calculation of Row Sums}
For the \(i\)-th row, the row sum is:
\begin{align} 
S_i = \sum_{j = 1}^{2^n} \left|o_{ij} - \frac{\text{tr}(O)}{2^n}\delta_{ij}\right| = \sum_{j\neq i} |o_{ij}| + \left|o_{ii} - \frac{\text{tr}(O)}{2^n}\right|
\end{align} 
\begin{enumerate}
    \item \(\sum_{j\neq i} |o_{ij}|\)\\
    \\
Each row of the matrix contains \( k \) non-zero elements, with the distribution of these non-zeros depending on the value of the diagonal element \( o_{ii} \): specifically, if \( o_{ii} \neq 0 \), there are \( k - 1 \) non-zero elements among the off-diagonal positions \( j \neq i \), whereas if \( o_{ii} = 0 \), all \( k \) non-zero elements lie in the off-diagonal positions \( j \neq i \). The probability that the diagonal element \( o_{ii} \neq 0 \) is given by \( \frac{k}{2^n} \), a value that becomes particularly small when \( k \ll 2^n \); in such cases, the overwhelming majority of diagonal elements \( o_{ii} \) will be zero, meaning the non-zero elements in most rows are concentrated entirely in the off-diagonal positions.\\
\\
    Let \(m_i\) be the number of non-zero elements in row \(i\) for \(j \neq i\), then \(m_i = k\) or \(k - 1\). Each non-zero \(o_{ij} \sim N(0, 1)\), so \(|o_{ij}|\) follows a half-normal distribution with expectation:
\begin{align} 
    \mathbb{E}[|o_{ij}|] = \sqrt{\frac{2}{\pi}} \approx 0.7979
\end{align} 
    and variance:
\begin{align} 
    \text{Var}(|o_{ij}|) = 1 - \frac{2}{\pi} \approx 0.3634
\end{align} 
    Thus:
\begin{align} 
    \sum_{j\neq i} |o_{ij}| \approx m_i \cdot \sqrt{\frac{2}{\pi}}
\end{align} 
    \item \(\left|o_{ii} - \frac{\text{tr}(O)}{2^n}\right|\)\\
    \\
       When \( o_{ii} \neq 0 \), the diagonal element \( o_{ii} \) follows a normal distribution \( N(0, 1) \); given that \( \frac{\text{tr}(O)}{2^n} \) is small, the absolute deviation \( \left| o_{ii} - \frac{\text{tr}(O)}{2^n} \right| \) can be approximated by \( |o_{ii}| \), which, for a standard normal variable, has an expected value of \( \sqrt{\frac{2}{\pi}} \). In contrast, when \( o_{ii} = 0 \), this deviation simplifies to \( \left| \frac{\text{tr}(O)}{2^n} \right| \), a quantity that remains very small due to the small magnitude of \( \frac{\text{tr}(O)}{2^n} \). Since the probability \( \frac{k}{2^n} \) (governing \( o_{ii} \neq 0 \)) is typically small in practice, the case \( o_{ii} = 0 \) dominates across most rows. As a result, for the majority of rows, the sum \( S_i \) (which reflects the relevant magnitude of elements in row \( i \)) can be approximated by the sum of absolute values of all non-zero off-diagonal elements in that row, i.e., \( S_i \approx \sum_{j: o_{ij} \neq 0} |o_{ij}| \), where this sum consists of exactly \( k \) terms corresponding to the non-zero entries in the row.\\
       \\
    \item Distribution of \(S_i\)\\
    \\
   Approximately, the sum \( S_i \approx \sum_{j: o_{ij} \neq 0} |o_{ij}| \) can be characterized as the sum of \( k \) independent half-normal random variables, each arising from the absolute value of a standard normal variable (consistent with the distribution of non-zero off-diagonal elements \( o_{ij} \)). For such a sum, the mean is given by \( k\sqrt{\frac{2}{\pi}} \), while the variance equals \( k\left(1 - \frac{2}{\pi}\right) \). As \( k \) becomes large, the central limit theorem applies, leading to a useful approximation where \( S_i \) follows a normal distribution: specifically, \( S_i \approx N\left(k\sqrt{\frac{2}{\pi}}, k\left(1 - \frac{2}{\pi}\right)\right) \). 
\end{enumerate}
Next, we estimate the maximum value.
\begin{align} 
\left\|O - \frac{\text{tr}(O)}{2^n}I\right\|_{\infty} = \max_{1\leq i\leq 2^n} S_i
\end{align} 
This is the maximum of \(2^n\) approximately normal random variables. According to extreme value theory, for \(m = 2^n\) random variables from \(N(\mu, \sigma^2)\), the maximum \(M_m\) behaves as:
\begin{align} 
M_m \approx \mu + \sigma\sqrt{2\log m}
\end{align} 
where:
\begin{itemize}
    \item \(\mu = k\sqrt{\frac{2}{\pi}}\)
    \item \(\sigma = \sqrt{k\left(1 - \frac{2}{\pi}\right)}\)
    \item \(m = 2^n\), so \(\log m = n\log 2 \)
\end{itemize}
Substituting, we get:
\begin{align} 
\left\|O - \frac{\text{tr}(O)}{2^n}I\right\|_{\infty} \approx k\sqrt{\frac{2}{\pi}} + \sqrt{k\left(1 - \frac{2}{\pi}\right)}\cdot\sqrt{2n\log 2}
\end{align} 

This norm is approximately \(O(k + \sqrt{kn})\).When \(k\) is fixed and \(n\) increases, the norm grows as \(\sqrt{n}\); When \(n\) is fixed and \(k\) increases, the dominant term is \(k\).

This estimation assumes that \(k\) is sufficiently large for the normal approximation and that for large \(n\), the effect of \(\frac{\text{tr}(O)}{2^n}\) is negligible. For very small \(k\) or \(n\), a more precise analysis may be needed, but this expression provides a reasonable scale for typical cases.\\
\\
Finally, the total number of measurements $T$ is
\begin{align} 
T = \left\lceil \frac{34 \cdot \text{max\_shadow\_norm} \cdot \left\lceil 2\log(2M/\delta) \right\rceil}{\varepsilon^2} \right\rceil
\end{align} 
Substituting:
\begin{align} 
T = \left\lceil \frac{34 \cdot \left[ k\sqrt{\frac{2}{\pi}} + \sqrt{k \left( 1 - \frac{2}{\pi} \right) \cdot 2n\log 2} \right]^2 \cdot \left\lceil 2\log(2M/\delta) \right\rceil}{\varepsilon^2} \right\rceil
\end{align} 
At this point, the part concerning the quantum resource estimation in Statement 3 is proven.
\subsection{Estimation of classical resources (LHM): proof of Statement 3(b)}
\label{sec:b4}
To estimate the number of floating-point operations in the expectation value calculation part, we analyze the floating-point operations in the core computational parts. The key components include:\\
\\
\textbf{Constructing Single-Qubit Density Matrices: }The ground state projection step involves no floating-point operations, as it relies solely on direct assignment to map the system to the ground state configuration, avoiding any numerical computation. In contrast, the basis transformation, which corresponds to the operation \(U^{\dagger}\rho U\), entails more involved numerical processing: specifically, it requires two \(2\times2\) complex matrix multiplications. Each of these matrix multiplications can be decomposed into real arithmetic operations, consisting of 8 real multiplications and 4 real additions, totaling 12 floating-point operations (FLOPs) per multiplication. Scaling this to the system size, the total computational cost for the basis transformation thus amounts to 2 multiplications multiplied by 12 FLOPs, resulting in 24 FLOPs per qubit, a cost that accumulates linearly with the number of qubits in the system.\\
\\
\textbf{Computing \(3\rho - I\): }Scalar multiplication (applied three times in the process) involves performing a single floating-point operation (FLOP) on each of the 4 elements of the matrix, resulting in a total of 4 FLOPs for this step; this operation scales the matrix by a scalar factor, adjusting the magnitude of its elements in a uniform manner. Following this, the matrix subtraction step (specifically subtracting the identity matrix \( I \)) requires 1 FLOP per element for the 4 elements involved, also contributing 4 FLOPs, this subtraction modifies each element of the matrix by removing the corresponding entry from the identity matrix, which is critical for isolating the target components of the matrix. Combining these two steps, the total computational cost for the process amounts to 8 FLOPs per qubit, a cost that remains consistent across individual qubits and scales linearly with the number of qubits in larger systems, making it a predictable and manageable component of the overall computational framework.\\
\\
\textbf{Kronecker Product Accumulation:} At step \( j \), the input matrix involved in the computation has a size of \( 2^j \times 2^j \), a dimension that grows exponentially with \( j \) and directly influences the computational complexity of subsequent operations. Correspondingly, the number of non-zero elements in this matrix is \( 4^j \). For each Kronecker product operation performed at this step, the computational cost is determined by the number of non-zero elements and the arithmetic required per element: specifically, each complex multiplication (a key component of the Kronecker product) entails 6 floating-point operations (FLOPs), and with 4 such operations needed per non-zero element, the total FLOPs for the Kronecker product amount to \( 4^j \times 4 \times 6 = 24 \times 4^j \). 
\begin{itemize}
    \item Total accumulation operations:
\begin{align}
    \sum_{j = 0}^{n - 1} 24 \times 4^j = 24 \times \frac{4^n - 1}{4 - 1} = 8(4^n - 1) \text{ FLOPs}
\end{align}
\end{itemize}
\textbf{Expectation Value Calculation (\(\text{Tr}(O\sigma_i)\)): }The observable \( O \) contains \( 2^n \times k \) non-zero elements, a count that reflects the combined influence of the system size (governed by \( n \), the number of qubits) and the sparsity structure (determined by \( k \), the average number of non-zeros per row). For each of these non-zero elements, two key operations are performed: a complex multiplication, which involves 6 floating-point operations (FLOPs) due to the need to handle both real and imaginary components separately, and a subsequent addition, which contributes 1 FLOP to accumulate the result into the overall sum. When aggregated across all non-zero elements, these operations yield a total computational cost of \( 7 \times 2^n k \) FLOPs per observable.\\
\\
\textbf{Total Floating-Point Operations Expression: }For \(T\) samples and \(M\) observables, the calculation is:
\begin{align}
\text{FLOPs} = T \left[ \underbrace{8(4^n - 1)}_{\text{Kronecker product}} + \underbrace{24n+8n}_{\text{single-qubit operations}} + \underbrace{7M \cdot 2^n k}_{\text{expectation value calculation}} \right]
\end{align}
We already know that
\(
T = \left\lceil \frac{34 \cdot \left[ k\sqrt{\frac{2}{\pi}} + \sqrt{k \left( 1 - \frac{2}{\pi} \right) \cdot 2n\log 2} \right]^2 \cdot \left\lceil 2\ln(2M/\delta) \right\rceil}{\varepsilon^2} \right\rceil
\)\\
Substituting \( T \) into the above equation yields
\begin{align}
\text{Total FLOPs} = \left\lceil \frac{34}{\varepsilon^2} \left[ k\sqrt{\frac{2}{\pi}} + \sqrt{2kn\log 2 \left( 1 - \frac{2}{\pi} \right)} \right]^2 \cdot \left\lceil 2\log(2M/\delta) \right\rceil \right\rceil \times \left[ 8(4^n - 1) + 32n + 7M \cdot 2^n k \right]
\end{align}
At this point, the part concerning the classical resource estimation in Statement 3 is proven.

\subsection{Estimation of resources of direct quantum measurement} (LHM)
\label{sec:b5}
Since \(O_m\) is a \(2^n \times 2^n\) Hermitian matrix, direct measurement implies independently executing the measurement process for each \(O_m\).

\textbf{Statistical Requirements for Measurement Times: }The overarching objective is to estimate the expectation values of all \( M \) observables with a specified additive error \( \epsilon \) and a failure probability \( \delta \), ensuring that the computed estimates remain within \( \epsilon \) of the true expectation values for all observables with a confidence level of \( 1 - \delta \). To contextualize this goal, it is important to note the nature of the measurement results: each individual measurement of an observable \( O_m \) yields one of its eigenvalues, which is bounded within the interval \( [-\|O_m\|, \|O_m\|] \). Here, \( \|O_m\| \) denotes the spectral norm of \( O_m \), defined as the largest absolute value among its eigenvalues, a quantity that characterizes the maximum possible magnitude of the observable's measurement outcomes and thus plays a critical role in determining the precision requirements for the estimation process.

According to the Hoeffding inequality, for a random variable \(X\) (measurement result) with values in \([a, b]\), to ensure that the probability of the sample mean deviating from the true expectation by less than \(\epsilon\) is at least \(1 - \delta'\) (where \(\delta'\) is the failure probability for a single observable), the required number of samples \(N\) is:
\begin{align}
N \geq \frac{(b - a)^2}{2\epsilon^2} \ln\left(\frac{2}{\delta'}\right)
\end{align}
For \(O_m\):
\(a = -\|O_m\|\), \(b = \|O_m\|\), so \(b - a = 2\|O_m\|\).\\
Substituting:
\begin{align}
N_m = \left\lceil\frac{(2\|O_m\|)^2}{2\epsilon^2} \ln\left(\frac{2}{\delta'}\right)\right\rceil = \left\lceil\frac{2\|O_m\|^2}{\epsilon^2} \ln\left(\frac{2}{\delta'}\right)\right\rceil
\end{align}

\textbf{Adjustment for Multiple Observables: }Since there are \(M\) observables, we want the probability that all \(M\) estimates simultaneously satisfy the error \(\epsilon\) to be at least \(1 - \delta\). According to the union bound, setting the failure probability for each observable to \(\delta/M\) ensures the total failure probability does not exceed \(\delta\). Therefore:
\(\delta' = \delta/M\).
\(\ln\left(\frac{2}{\delta'}\right) = \ln\left(\frac{2M}{\delta}\right)\).
The number of measurements for each observable \(O_m\):
\begin{align}
N_m = \left\lceil\frac{2\|O_m\|^2}{\epsilon^2} \ln\left(\frac{2M}{\delta}\right)\right\rceil
\end{align}

\textbf{Total Number of Measurements: }Assuming that the \(M\) observables do not share measurements (i.e., not considering whether the observables can be diagonalized simultaneously), the total number of measurements is:
\begin{align}
N_{\text{total}} = \sum_{m = 1}^{M} N_m = \sum_{m = 1}^{M} \left\lceil\frac{2\|O_m\|^2}{\epsilon^2} \ln\left(\frac{2M}{\delta}\right)\right\rceil
\end{align}

\textbf{Calculation of the Spectral Norm \(\|O_m\|\): }
\( O_m \) is a sparse Hermitian matrix where each row contains \( k \) non - zero elements, and these non - zero entries follow the standard normal distribution \( \mathcal{N}(0, 1) \). For such matrices, to determine the spectral norm \( \| O_m \| \) (the maximum absolute eigenvalue, crucial for bounding measurement outcomes as discussed prior), numerical techniques like \( \texttt{scipy.sparse.linalg.eigsh} \) are employed to compute the largest and smallest eigenvalues of each \( O_m \); the maximum absolute value among these eigenvalues is then taken as \( \| O_m \| \). However, directly computing this spectral norm (equivalent to the largest singular value for Hermitian matrices) is computationally burdensome, especially for high - dimensional sparse matrices. Thus, an upper - bound approximation is utilized: \( \| O_m \| \leq \sqrt{\| O_m \|_1 \cdot \| O_m \|_\infty} \), where \( \| O_m \|_1 \) denotes the matrix 1 - norm (the maximum absolute column sum) and \( \| O_m \|_\infty \) is the matrix infinity - norm.\\
\\
For Hermitian matrices, \(\| O_m \|_1 = \| O_m \|_\infty\), because the row and column sums are symmetric. We can calculate the 1-norm (\( \text{norm}_1 \)) and the infinity norm (\( \text{norm}_{\infty} \)) of the matrix through a program, and then take:
\begin{align}
\text{spectral\_bound} = \sqrt{\text{norm\_1} \cdot \text{norm\_inf}}
\end{align}
and substitutes this upper bound into the calculation formula for \(N_m\).\\

\textbf{Analysis of relationship Between Spectral Norm and \(k\): }
We first conduct an analysis of the variance of a matrix as the initial step of our research or calculation process.
\begin{itemize}
    \item \textbf{Diagonal}: Real numbers drawn from the standard normal distribution \(\mathcal{N}(0, 1)\).  \( O_{ii} \sim \mathcal{N}(0, 1) \), so the variance \( \mathbb{E}[O_{ii}^2] = 1 \).
    \item \textbf{Off-diagonal}: \( O_{ij} = a + bi \ (i \neq j) \), where \( a, b \sim \mathcal{N}(0, 1) \) are independent. Then:
\begin{align}
    \mathbb{E}[|O_{ij}|^2] = \mathbb{E}[a^2 + b^2] = \mathbb{E}[a^2] + \mathbb{E}[b^2] = 1 + 1 = 2
\end{align}
\end{itemize}

For a sparse random matrix, the spectral norm \( \|O_m\| \) mainly depends on the off - diagonal elements. Because when \( k \) is large, the contribution of the diagonal elements can be neglected (their variance is 1, while the variance of the off - diagonal elements is 2, and the quantity is \( k - 1\approx k \)). On the whole, the average variance of each non - zero element is:

\begin{align}
\text{Average variance} = \frac{1\cdot\mathbb{E}[O_{ii}^2]+(k - 1)\cdot\mathbb{E}[|O_{ij}|^2]}{k}=\frac{1\cdot1+(k - 1)\cdot2}{k}=2-\frac{1}{k}\approx 2\quad\text{when}\ k\gg 1
\end{align}
Let \( v^2 = 2 \) represent the variance of the off - diagonal elements.\\
\\
Following the initial step, we perform the estimation of the spectral norm in Random Matrix Theory as the second part of our methodology.\\
In random matrix theory, for a sparse Hermitian matrix, each row has \( k \) non - zero elements, the element mean is 0, and the variance is \( v^2 \). The asymptotic behavior of the spectral norm \( \|O_m\| \) is given by the following formula(this formula can be found in \cite{benaych2020spectral}):

\begin{align}
\|O_m\|\approx 2v\sqrt{d}
\end{align}
where \( d \) is the average degree of the graph (that is, the number of off - diagonal elements connected to each vertex). Here:\\
The adjacency graph of the matrix is a random graph, where each row contains \( k \) non - zero elements (encompassing the diagonal element). Specifically, the diagonal elements correspond to self - loops within the graph structure, while the off - diagonal elements represent edges between different vertices. Given that each row has \( k \) non - zero entries, with one being the diagonal element and \( k - 1 \) being off - diagonal, the degree of each vertex (accounting for self - loops) is \( k \). However, when considering only off - diagonal connections (as self - loops do not contribute to the edges between distinct vertices), the average degree of these off - diagonal edges is \( d = k - 1 \). 
\\
\\
For large \( n \) (matrix dimension \( N = 2^n\rightarrow\infty \)) and \( k \) that is fixed or grows moderately, the spectral norm is mainly dominated by the off - diagonal part. Standard results  show that:

\begin{align}
\|O_m\|\approx 2v\sqrt{d},\quad\text{where}\ d = k - 1
\end{align}
Substitute the variance \( v^2 = 2 \):

\begin{align}
\|O_m\|\approx 2\sqrt{2}\sqrt{k - 1}
\end{align}
When \( k\gg 1 \), \( \sqrt{k - 1}\approx\sqrt{k} \). Therefore:

\begin{align}
\|O_m\|\approx 2\sqrt{2}\sqrt{k}
\end{align}
\\
\noindent Substitute into the Measurement Count Formula: \\
Substitute \( \| O_m \| \approx 2\sqrt{2}\sqrt{k} \) into \( N_m \):  
\begin{align}
N_m \approx \frac{2(2\sqrt{2}\sqrt{k})^2 \ln\left( \frac{2M}{\delta} \right)}{\epsilon^2} = \frac{16k \ln\left( \frac{2M}{\delta} \right)}{\epsilon^2}
\end{align} 
Assuming that the spectral norms of all \( M \) observables are approximately equal, the total number of measurements is:  
\begin{align}
N_{\text{total}} \approx M \cdot \frac{16k \ln\left( \frac{2M}{\delta} \right)}{\epsilon^2}
\end{align}
With this, the proof of Statement 4 is complete.

\section{Numerical experiments}\label{secD}
Building on the two observable classes (LCP and LHM) introduced in the main text, both of which have found broad applications in physical experiments, our resource-comparison framework for selecting between classical shadows and direct quantum measurement is expected to be applicable to a wide range of practical settings. In what follows, we present a set of numerical experiments illustrating how the resource costs of these two approaches vary with the relevant problem parameters for the two observable families.

We first focus on large Hermitian matrix (LHM) observables. In many-body physics, quantum chemistry, and quantum chaos experiments, many physically natural observables (e.g., swap operators, projection operators, or global operators evolved in time) appear, in the computational basis, as Hermitian matrices of exponentially large dimension but with structured sparsity. Although such operators can in principle be expanded as linear combinations of Pauli tensor products, the number of Pauli terms typically grows exponentially, making the Pauli representation unnatural in certain experimental and numerical settings. Hence, treating them directly as LHM observables better reflects realistic measurement workflows.
\subsection{Measuring the second R\'enyi entropy}
\begin{figure*}
    \centering
    \begin{subfigure}{0.48\textwidth}
        \centering
        \includegraphics[width=\textwidth, trim=10pt 10pt 10pt 0pt, clip]{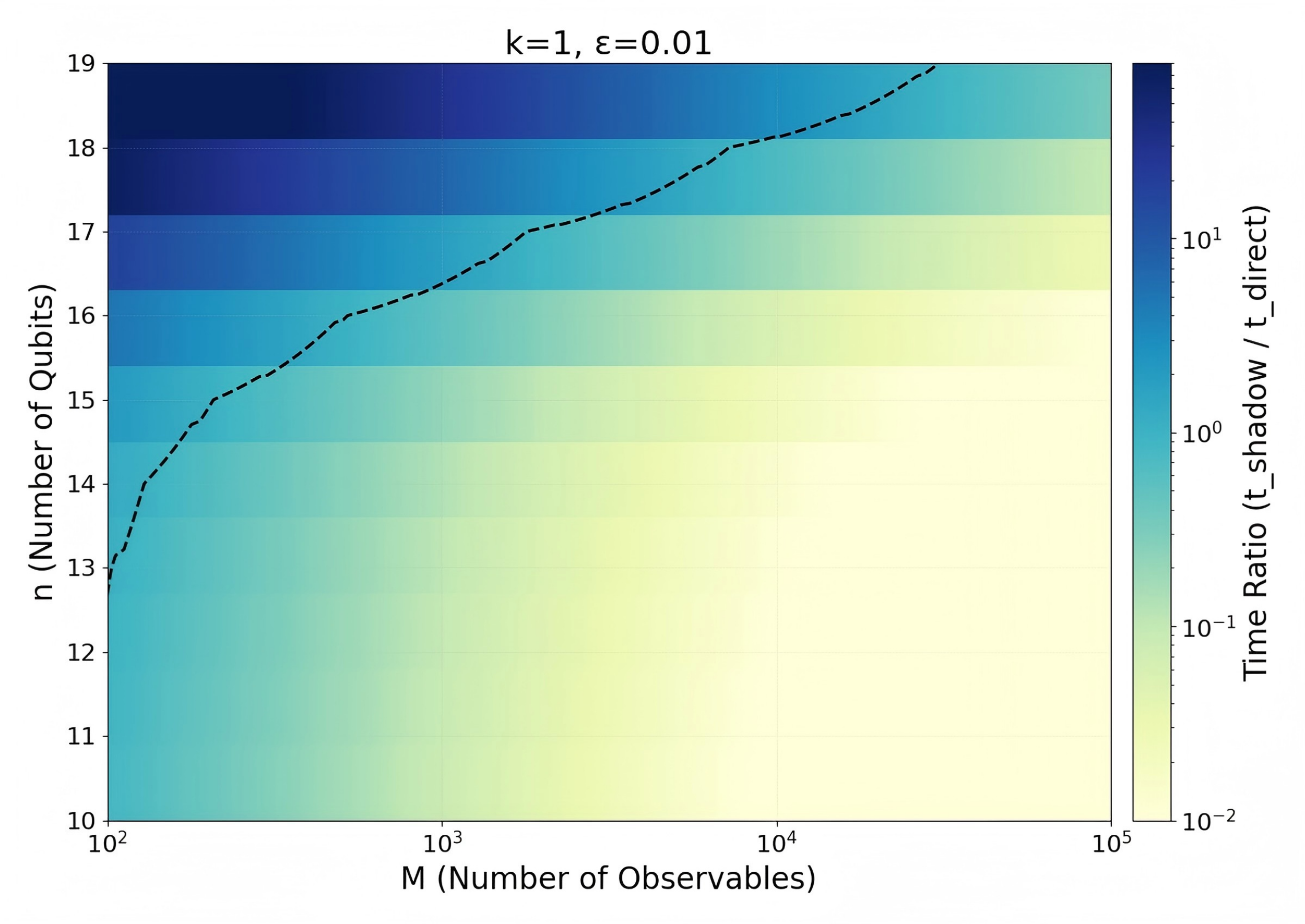}
        \caption{Runtime Ratio.}
        \label{lhmsubfi:a}
    \end{subfigure}
    \begin{subfigure}{0.48\textwidth}
        \centering
        \includegraphics[width=\textwidth, trim=10pt 10pt 10pt 0pt, clip]{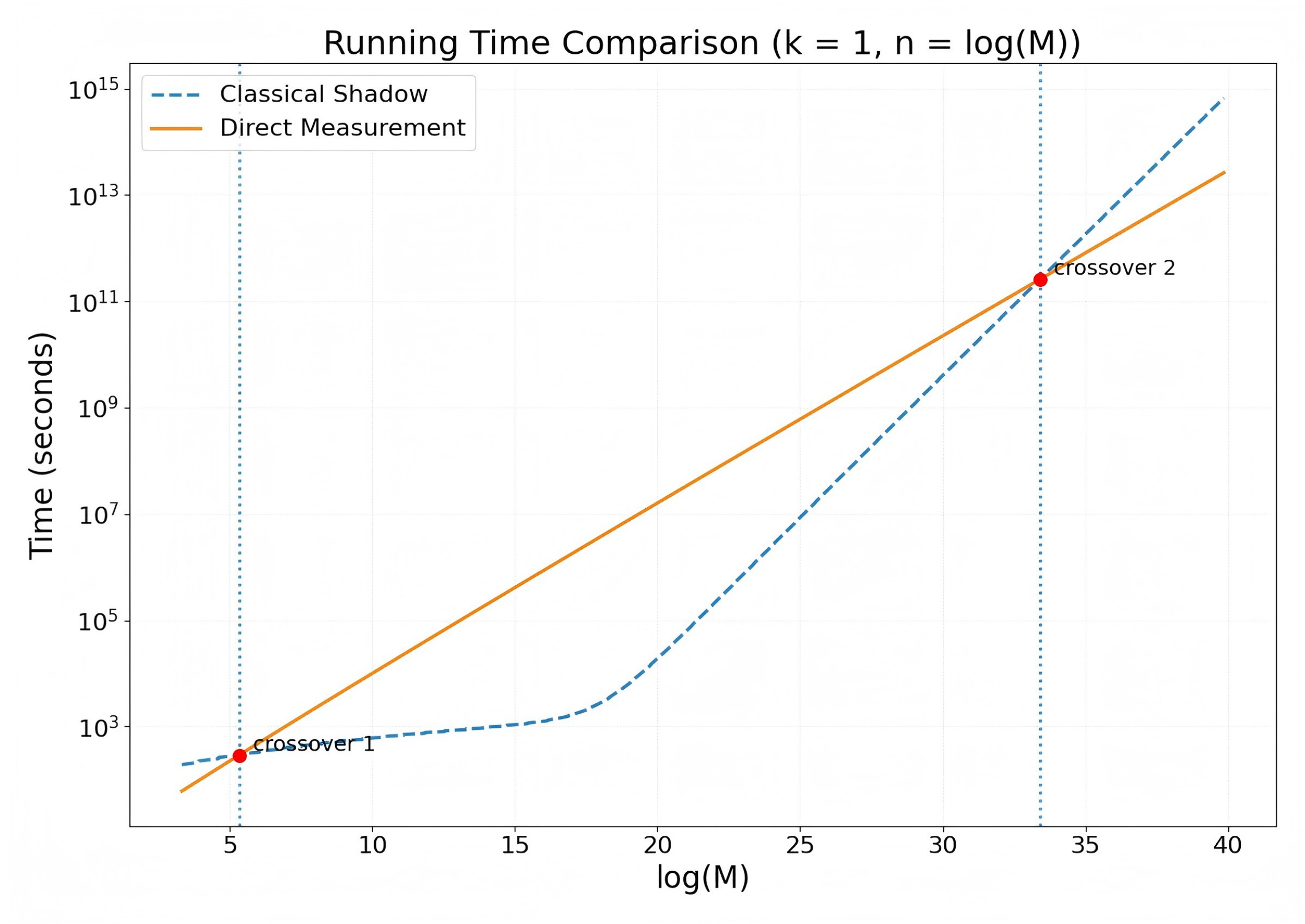}
        \caption{Runtime Comparison.}
        \label{lhmsubfi:b}
    \end{subfigure}

        \caption{
    \textbf{Comparison of Classical Shadow Method and Direct Quantum Measurement Method in measuring the second R\'enyi entropy.} In the plots, $M$ represents the number of observables, $n$ is the number of qubits, $k$ is the sparsity,  $\epsilon$ is the precision tolerance of the prediction. In the experiments measuring the second R\'enyi entropy, \( k = 1 \) is kept fixed.
    (a): Just like the heatmap in the main text, the black dashed line indicates the threshold for resource balance between the two methods. Above the dashed line, the direct quantum measurement approach outperforms the classical shadow method, while below it, the opposite holds. (b): Runtime comparison between classical shadow method and direct quantum measurement method for $S_{A_i}$, assuming the number of qubits $n$ as $\log (M)$.
    }
    \label{fig:lhmbig_figure}
\end{figure*}

In quantum many-body and quantum simulation experiments, one often aims to characterize the entanglement structure of subsystems. Suppose the system state $\rho$ is prepared under a Hamiltonian $H$ (e.g., a local spin model $H=\sum_i h_i Z_i+\sum_{i<j}J_{ij}Z_iZ_j$), either as a ground state or via unitary time evolution $\rho(t)=e^{-iHt}\rho_0 e^{iHt}$. To access entanglement information for multiple subsystems, consider a family of subsystems $\mathcal{A}=\{A_1,\dots,A_M\}$, and define the reduced state for each subsystem as $\rho_{A_i}=\Tr_{\overline{A_i}}(\rho)$. The second-order R\'enyi entropy\cite{PhysRevLett.104.157201,Brydges_2019} is then
\begin{align}
S_2(A_i)=-\log \Tr(\rho_{A_i}^2),\qquad i=1,\dots,M.
\end{align}
Experimentally, one may introduce two independent copies $\rho\otimes\rho$ and use the identity
\begin{align}
\Tr(\rho_{A_i}^2)=\Tr[(\rho\otimes\rho)\,S_{A_i}],
\end{align}
which expresses each target quantity as the expectation value of an observable, where $S_{A_i}$ is the swap operator acting on subsystem $A_i$ across the two copies, defined by
$S_{A_i}\ket{x}\ket{y}=\ket{y}\ket{x}$ (with $x,y\in\{0,1\}^{n_{A_i}}$, and $n_{A_i}$ denoting the number of qubits in subsystem $A_i$).
Accordingly, an LHM observable family for multi-subsystem R\'enyi-entropy measurements can be chosen as $\{O_i\}_{i=1}^M$ with $O_i=S_{A_i}$.
In the computational basis of two copies, each $S_{A_i}$ is a Hermitian matrix of dimension $2^{2n_{A_i}}\times 2^{2n_{A_i}}$, and each row contains exactly one nonzero entry, corresponding to sparsity $k=1$ (which may be viewed as a special case of our LHM parameterization with effective qubit number $n=2n_{A_i}$ and sparsity $k=1$). Although the Pauli decomposition of $S_{A_i}$ contains exponentially many terms, its expectation value can be accessed experimentally via swap tests or randomized measurement protocols (e.g., randomized measurements / classical shadows) without explicitly expanding it into Pauli form; this has been demonstrated in multiple experimental implementations~\cite{PhysRevA.98.052334,doi:10.1126/science.aau4963}. Concretely, a swap test introduces an ancilla qubit and implements a controlled-SWAP operation on subsystem $A_i$. The statistical average of the ancilla $Z$-measurement outcomes yields $\Tr[(\rho\otimes\rho)S_{A_i}]$. A comparison of the resources required by the two methods for measuring the second R\'enyi entropy is presented in Figure~\ref{fig:lhmbig_figure}. 

Figure~\ref{lhmsubfi:a} displays a heatmap of the boundary between the two methods, analogous to those in the main text. It provides an intuitive visualization of the performance boundary between the classical shadow method and the direct quantum measurement approach in such experiments, as well as indicating which method is optimal for specific regimes of \( M \) and \( n \). Figure~\ref{lhmsubfi:b} plots the resource consumption of both methods as a function of \( M \), under the assumption that \( n = \log(M) \). From the intersection point in Figure~\ref{lhmsubfi:b}, we observe that, compared to the case of \( k = \log M \) discussed in the main text, the regime in which the classical shadow method exhibits a clear advantage has expanded significantly.

In practical scenarios, however, the SWAP test for measuring the second-order R\'enyi entropy incurs substantial quantum gate costs, with circuit depth constituting a significant and non-negligible portion of the resource consumption. This implies that resource estimates for the direct quantum measurement approach are frequently underestimated in this context. As a result, the regime where the classical shadow method holds an advantage is expected to appear earlier and vanish later as $M$ grows.

\subsection{$k$-body correlation observables in the one-dimensional Heisenberg XXZ chain}

In quantum many-body and quantum simulation experiments, a central goal is to characterize higher-order correlation and entanglement properties of quantum states prepared in interacting spin systems. We consider an $n$-qubit system governed by the one-dimensional Heisenberg XXZ Hamiltonian\cite{PhysRevLett.105.095702,Sato_2007}
\begin{equation}
H
=
\sum_{i=1}^{n-1}
\left(
J_x X_i X_{i+1}
+
J_y Y_i Y_{i+1}
+
J_z Z_i Z_{i+1}
\right),
\end{equation}
where $X_i, Y_i, Z_i$ denote Pauli operators acting on site $i$, and $J_x, J_y, J_z$ are anisotropic coupling strengths. The system state $\rho$ may be prepared either as a ground state of $H$ or via unitary time evolution,
\begin{equation}
\rho(t)=e^{-iHt}\rho_0 e^{iHt},
\end{equation}
starting from an initial state $\rho_0$.

To probe genuine multi-body correlations beyond two-point functions, we focus on the complete set of $k$-body correlation observables. Specifically, we consider the family of all $k$-site subsets
\begin{equation}
\mathcal{A}_k
=
\left\{
A_1,\dots,A_{|\mathcal{A}_k|}
\right\},
\qquad
A_i=\{i_1,i_2,\dots,i_k\},
\end{equation}
with $1\le i_1<i_2<\cdots<i_k\le n$, and define for each subset the corresponding $k$-body correlators
\begin{equation}
\bigl\langle
\sigma_{i_1}^{a_1}
\sigma_{i_2}^{a_2}
\cdots
\sigma_{i_k}^{a_k}
\bigr\rangle
=
\Tr\!\left[
\rho\,
\sigma_{i_1}^{a_1}
\otimes
\sigma_{i_2}^{a_2}
\otimes
\cdots
\otimes
\sigma_{i_k}^{a_k}
\right],
\end{equation}
where $a_j\in\{x,y,z\}$.

Accordingly, the observable family of interest can be written as
\begin{equation}
\mathcal{O}_k
=
\left\{
O_i
\right\}_{i=1}^{M},
\qquad
O_i
=
\sigma_{i_1}^{a_1}
\otimes
\sigma_{i_2}^{a_2}
\otimes
\cdots
\otimes
\sigma_{i_k}^{a_k},
\end{equation}
which consists of Pauli strings acting nontrivially on $k$ distinct sites. For the \emph{full set} of $k$-body correlators, the total number of target observables is
\begin{equation}
M
=
\binom{n}{k}\,3^k,
\end{equation}
which grows rapidly with $n$ and $k$, rendering brute-force measurement strategies based on individual Pauli measurements impractical even for moderate system sizes.

\begin{figure*}
    \centering
    \begin{subfigure}{0.54\textwidth}
        \centering
        \includegraphics[width=\textwidth, trim=10pt 10pt 10pt 0pt, clip]{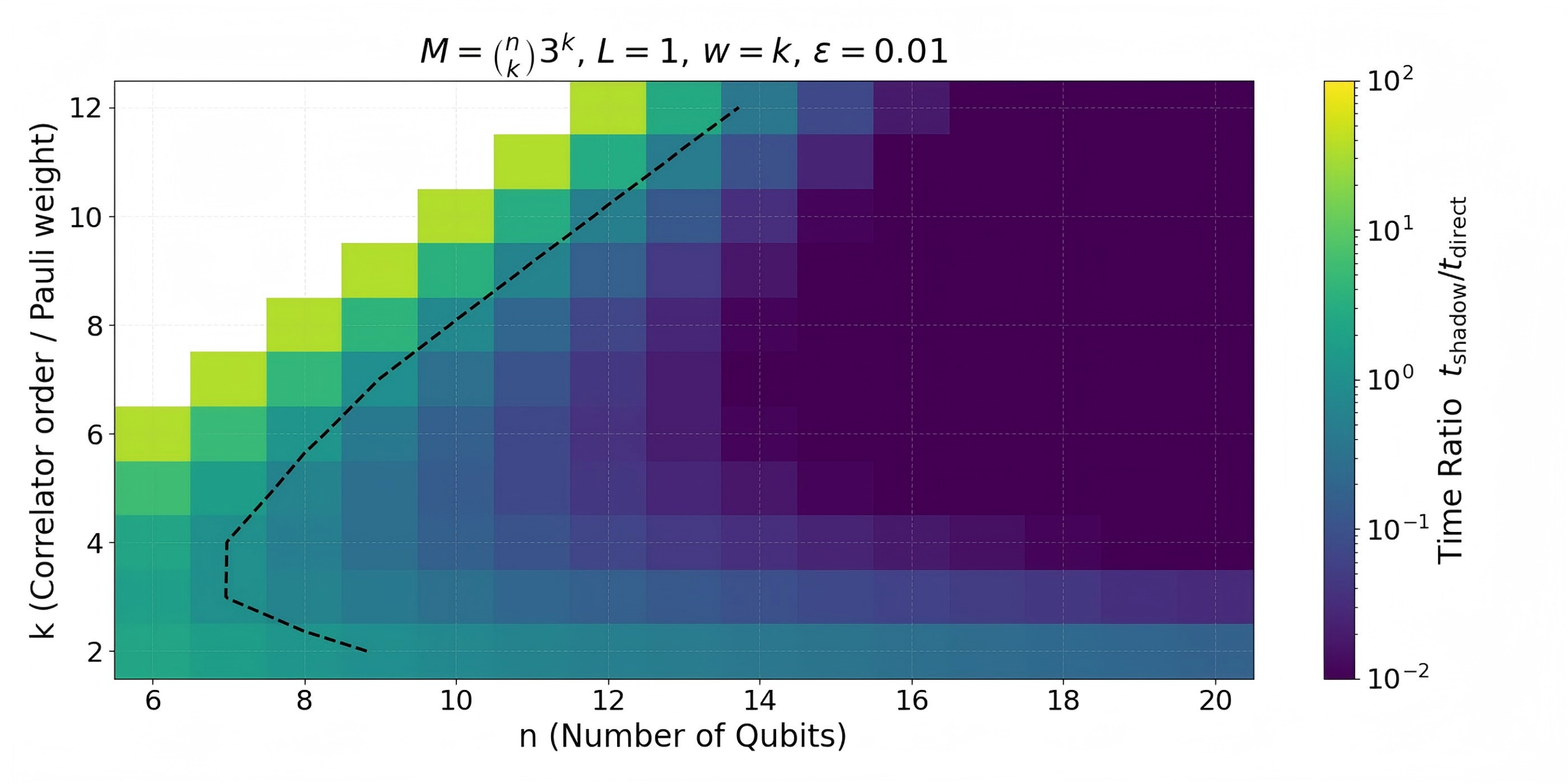}
        \caption{Runtime Ratio.}
        \label{lcpsubfi:a}
    \end{subfigure}
    \begin{subfigure}{0.44\textwidth}
        \centering
        \includegraphics[width=\textwidth, trim=10pt 10pt 10pt 0pt, clip]{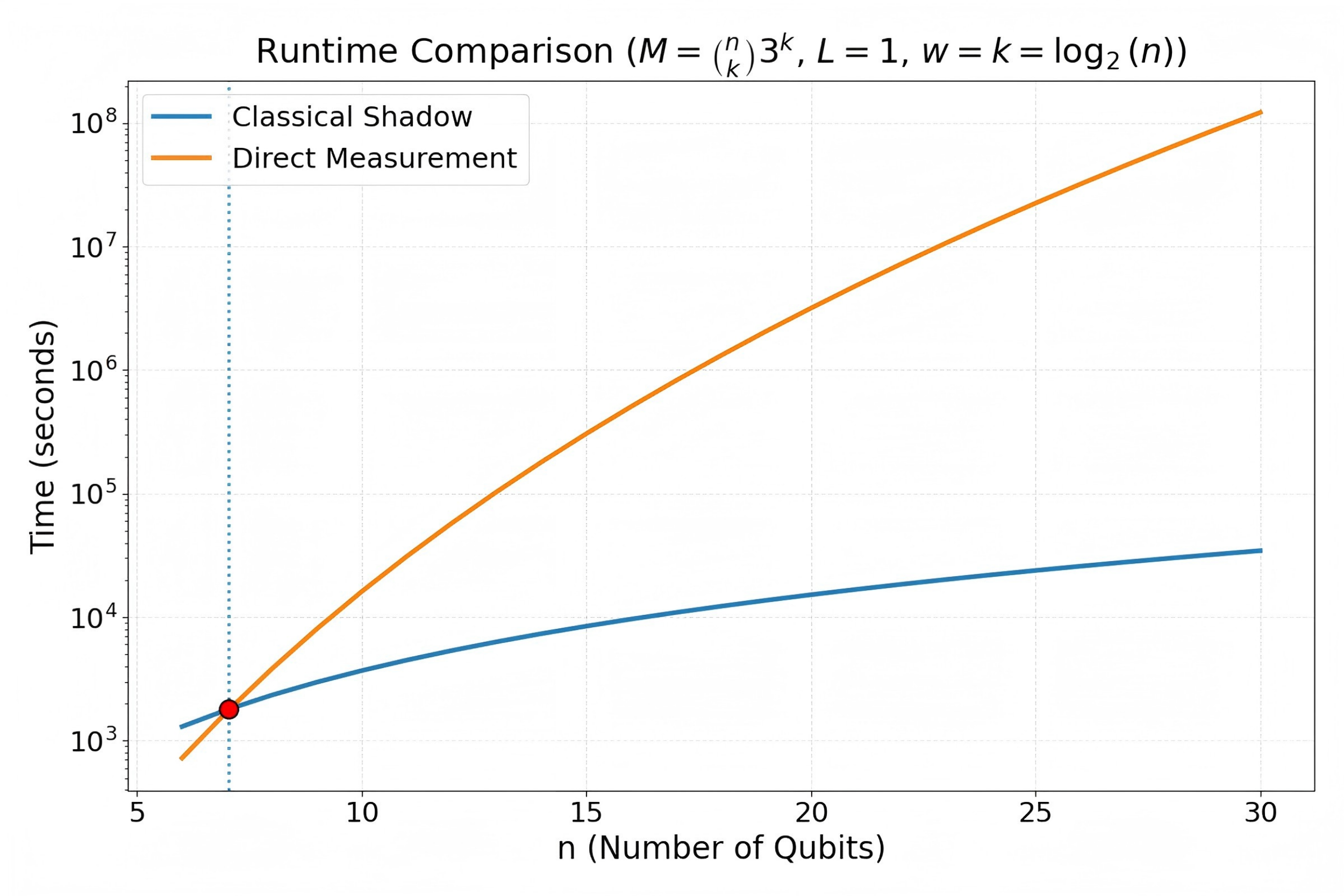}
        \caption{Runtime Comparison.}
        \label{lcpsubfi:b}
    \end{subfigure}

        \caption{
  \textbf{Comparison of Classical Shadow Method and Direct Quantum Measurement Method in measuring $k$-body correlation observables in the one-dimensional Heisenberg XXZ chain.} In the plots, $M$ represents the number of observables, $n$ is the number of qubits, $w$ and $k$ both represent Pauli weight, $L$ is the number of terms per observables, $\epsilon$ is the precision tolerance of the prediction. In the experiments measuring $k$-body correlation observables in the one-dimensional Heisenberg XXZ chain, \( L=1 \) remains constant, and the relationship \( M = \binom{n}{k} \, 3^k \) holds.
    (a): Just like the heatmap in the main text, the black dashed line indicates the threshold for resource balance between the two methods. To the left of the dashed line, the direct quantum measurement approach outperforms the classical shadow method, while to the right of it, the opposite holds. (b): Runtime comparison between classical shadow method and direct quantum measurement method for $\sigma_{i_1}^{a_1}
\sigma_{i_2}^{a_2}
\cdots
\sigma_{i_k}^{a_k}$, assuming the Pauli weight $w$ or $k$ as $\log_2(n)$.
    }
    \label{fig:lcpbig_figure}
\end{figure*}

Each observable $O_i$ is Hermitian with eigenvalues $\pm 1$ and admits a sparse representation as a single Pauli string on the full $n$-qubit Hilbert space. Although the number of target observables is large, all of them lie in the same Pauli operator basis, making them naturally amenable to randomized measurement protocols. In particular, classical shadow tomography enables the simultaneous estimation of all $k$-body correlators in $\mathcal{O}_k$ using a number of measurement settings that scales only logarithmically with $M$, thereby providing a scalable approach to accessing high-order correlation functions in quantum many-body systems. A comparison of the resources required by the two methods for measuring measuring $k$-body correlation observables in the one-dimensional Heisenberg XXZ chain is presented in Figure~\ref{fig:lcpbig_figure}. 

In the heatmap of Figure~\ref{lcpsubfi:a}, we take \( n \) and \( k \) as the horizontal and vertical axes, respectively. Since here \( M = \binom{n}{k} \, 3^k \), \( M \) is uniquely determined by \( n \) and \( k \). Consequently, a heatmap with \( M \) versus \( k \) or \( M \) versus \( n \) would only produce a vertical or horizontal boundary line, which is essentially meaningless. Therefore, we consider only the boundary plot in the \( n \)-\( k \) plane. The blank region in the upper-left corner corresponds to missing entries where \( k > n \), as the Pauli weight clearly cannot exceed the total number of qubits. From this plot, one can visually discern which method is advantageous for fixed parameters \( n \), \( w \), and \( L \), thereby significantly saving resources and improving the efficiency of method selection. The curve plot in Figure~\ref{lcpsubfi:b} is similar to the previous ones, except that the independent variable is changed to the qubit number \( n \), with the Pauli weight assumed to be \( \log_2(n) \).

We now consider a special case of direct quantum measurement for $k$-body Pauli observables, where the number of required measurement settings can be substantially reduced. As a simple example, let $w=k=2$, and suppose we aim to estimate the expectation values of all nearest-neighbor two-body Pauli observables,
\begin{align}
\{\,X_iX_{i+1},\, Y_iY_{i+1},\, Z_iZ_{i+1}\;:\;0\le i\le n-1\,\},
\end{align}
where we assume periodic boundary conditions for the index $i+1$ (taken modulo $n$) when necessary. In the standard setting one may take $M=3n$ measurement settings, corresponding to measuring each term separately. However, in this particular case we may adopt the idea of \emph{overlapping tomography}~\cite{PhysRevLett.124.100401} and perform only three global measurement settings: measuring all qubits in the $X$ basis, all qubits in the $Y$ basis, and all qubits in the $Z$ basis. This yields the global correlators
\begin{align}
\mathrm{tr}\!\left(X_1\otimes X_2\otimes\cdots\otimes X_n\,\rho\right),\quad
\mathrm{tr}\!\left(Y_1\otimes Y_2\otimes\cdots\otimes Y_n\,\rho\right),\quad
\mathrm{tr}\!\left(Z_1\otimes Z_2\otimes\cdots\otimes Z_n\,\rho\right),
\end{align}
from which one can obtain the local nearest-neighbor correlators
\begin{align}
\big\{\,
\mathrm{tr}\!\left(X_i\otimes X_{i+1}\,\rho\right),\ 
\mathrm{tr}\!\left(Y_i\otimes Y_{i+1}\,\rho\right),\ 
\mathrm{tr}\!\left(Z_i\otimes Z_{i+1}\,\rho\right)
\;:\;0\le i\le n-1
\,\big\}
\end{align}
by taking appropriate marginals (equivalently, partial traces) over the remaining subsystems. In this way, the number of measurement settings can be reduced from $M=3n$ to $M=3$, at the expense of an additional classical post-processing step associated with marginalization/partial trace. We next provide a simple resource estimate for this trade-off.

Without overlapping measurements, for $(M=3n,\ L=1,\ \epsilon=\delta=0.01)$, Statement~2 implies that the required number of measurement shots is
\begin{align}
\frac{0.5\, M L^3}{\epsilon^2}\log\!\left(\frac{2ML}{\delta}\right)
=15000\,n\,\log(600n).
\end{align}
With overlapping measurements, the number of settings is $M=3$ (while $L=1$ and $\epsilon=\delta=0.01$ remain unchanged), and the required number of shots becomes
\begin{align}
\frac{0.5\, M L^3}{\epsilon^2}\log\!\left(\frac{2ML}{\delta}\right)
=15000\,\log(600)\approx 95953.94.
\end{align}

We further account for the cost of computing the required marginals/partial traces, and quantify the classical post-processing cost in terms of the total number of floating-point operations (flops), as derived below.

For each global measurement basis $\alpha\in\{X,Y,Z\}$, we repeat the measurement $N_\alpha$ times. The $m$-th shot yields an outcome vector
\begin{align}
\mathbf{s}^{(\alpha,m)}=(s^{(\alpha,m)}_0,\dots,s^{(\alpha,m)}_{n-1}),\qquad 
s^{(\alpha,m)}_j\in\{+1,-1\},
\end{align}
where we represent single-qubit eigenvalues by $\pm1$ (equivalently, one may use $\{0,1\}$ outcomes via a fixed mapping). The global measurement in basis $\alpha$ induces an empirical joint distribution $p^{(\alpha)}(\mathbf{s})$ over $\mathbf{s}$. The two-site marginal for sites $(i,i+1)$, obtained by summing over all remaining coordinates (i.e., taking a partial trace / marginalization), is
\begin{align}
p^{(\alpha)}_{i,i+1}(a,b)
=\sum_{\mathbf{s}:\,s_i=a,\,s_{i+1}=b} p^{(\alpha)}(\mathbf{s}),\qquad a,b\in\{\pm1\}.
\end{align}
The target two-body Pauli expectation value can then be written, e.g. for $\alpha=X$, as
\begin{align}
\langle X_i X_{i+1}\rangle=\sum_{a,b\in\{\pm1\}} ab \, p^{(X)}_{i,i+1}(a,b).
\end{align}
Importantly, this quantity can be computed directly from the global outcomes without explicitly forming the marginals:
\begin{align}
\langle \alpha_i \alpha_{i+1}\rangle
=\mathbb{E}_{\mathbf{s}\sim p^{(\alpha)}}[s_i s_{i+1}]
\approx
\widehat{E}^{(\alpha)}_i
:=\frac{1}{N_\alpha}\sum_{m=1}^{N_\alpha} s^{(\alpha,m)}_i\, s^{(\alpha,m)}_{i+1}.
\end{align}
Consequently, the marginalization cost can be implemented as a streaming accumulation. For a fixed $\alpha$, we maintain accumulators $A^{(\alpha)}_i\leftarrow 0$ for all $i=0,1,\dots,n-1$ (with periodic boundary conditions so that $i+1$ is taken modulo $n$), and upon receiving shot $m$ we update
\begin{align}
A^{(\alpha)}_i \leftarrow A^{(\alpha)}_i + s^{(\alpha,m)}_i s^{(\alpha,m)}_{i+1},\qquad \forall i.
\end{align}
For each $i$, this update uses one floating-point multiplication and one floating-point addition, hence each shot costs $2n$ flops to update all nearest-neighbor correlators. After $N_\alpha$ shots, we output $\widehat{E}^{(\alpha)}_i=A^{(\alpha)}_i/N_\alpha$, which requires $n$ floating-point divisions. Therefore, the total post-processing cost for basis $\alpha$ is
\begin{align}
F_\alpha=(2n)N_\alpha+n = n(2N_\alpha+1).
\end{align}
Summing over $\alpha\in\{X,Y,Z\}$ yields
\begin{align}
F_{\mathrm{tot}}=\sum_{\alpha}F_\alpha
= n\Big(2(N_X+N_Y+N_Z)+3\Big).
\end{align}
In the common case $N_X=N_Y=N_Z=:N$, one obtains $F_{\mathrm{tot}}=3n(2N+1)=n(6N+3)$, or equivalently, in terms of the total number of shots $N_{\mathrm{tot}}:=N_X+N_Y+N_Z=3N$,
\begin{align}
F_{\mathrm{tot}}=n(2N_{\mathrm{tot}}+3).
\end{align}
With $\epsilon=\delta=0.01$ and three global settings, the sample complexity used above gives
\begin{align}
N_{\mathrm{tot}}=\frac{0.5\cdot 3\cdot 1^3}{\epsilon^2}\log\!\Big(\frac{2\cdot 3\cdot 1}{\delta}\Big)
=15000\log(600),
\end{align}
and thus
\begin{align}
F_{\mathrm{tot}}
=n\Big(2\cdot 15000\log(600)+3\Big)
= n\Big(30000\log(600)+3\Big)
\approx 1.9191\times 10^5\, n.
\end{align}
We continue to use time as the resource metric. The resources required by the direct quantum measurement method are given by
\begin{align}
1.9191\times 10^5\, n\times 10^{-15}+15000\log(600)\times10^{-8}.
\end{align}

This quantity is orders of magnitude smaller than that of the classical shadow method. Although the number of observables $M$ in this approach has not yet reached the regime where classical shadows demonstrate their advantage, the idea of overlapping measurements is frequently employed in practical experimental settings. For the general case of $k$-body Pauli observables, one can design measurement settings using covering arrays to achieve optimal performance~\cite{t6qb-kdcp}. We do not elaborate further on this point here.

However, it is worth noting that for the special case mentioned above, where \( M \) is relatively small and each term has a simple form, the direct quantum measurement approach invariably demonstrates an advantage. In contrast, when \( M \) becomes complex and numerous, the number of measurement settings required for global measurements increases accordingly. Although one can still reduce the number of settings by constructing covering arrays, the number of qubits that need to be manipulated in a single measurement grows with the system size \( n \). Such a space-for-time trade-off may not be reasonable under certain specific conditions. Therefore, in practical experiments, the choice of method should be determined on a case-by-case basis.


\section{Other supplementary materials}\label{secC}
\subsection{How to Decompose a \(2^n \times 2^n\) Hermitian Matrix into a Pauli Linear Combination}
We will start by introducing some fundamental concepts.
\begin{itemize}
    \item \textbf{Hermitian Matrix}: A matrix \(A\) is Hermitian if it satisfies \(A = A^{\dagger}\) (i.e., it is equal to its conjugate transpose). This means that for the matrix elements, \(a_{ij} = a_{ji}^*\), where \(*\) denotes complex conjugation. Hermitian matrices have real eigenvalues and are often used to represent observables in quantum mechanics.
    \\
    \item \textbf{Pauli Matrices}: Pauli matrices are a set of \(2 \times 2\) Hermitian and unitary matrices commonly used to describe quantum systems. They include:

        \(\sigma_x = \begin{bmatrix}0 & 1 \\ 1 & 0\end{bmatrix}\),
        \(\sigma_y = \begin{bmatrix}0 & -i \\ i & 0\end{bmatrix}\),
        \(\sigma_z = \begin{bmatrix}1 & 0 \\ 0 & -1\end{bmatrix}\),
    The identity matrix \(I = \begin{bmatrix}1 & 0 \\ 0 & 1\end{bmatrix}\) (although not strictly a Pauli matrix, it is often included in decompositions).

        In multi-qubit systems, higher-dimensional matrices are constructed through the tensor products of these matrices.
        \\
    \item \textbf{Decomposition Goal}: We need to prove that any \(2^n \times 2^n\) Hermitian matrix \(A\) (corresponding to \(n\) qubits) can be decomposed into a linear combination of tensor products of Pauli matrices:
\begin{align}
        A = \sum_{\mathbf{k}} c_{\mathbf{k}} (P_{k_1} \otimes P_{k_2} \otimes \cdots \otimes P_{k_n})
\end{align}
        where \(P_{k_i} \in \{I, \sigma_x, \sigma_y, \sigma_z\}\), and \(c_{\mathbf{k}}\) are real coefficients.
\end{itemize}
Next, we will now provide a detailed and rigorous proof of decomposability. To prove this, we need to show that the tensor products of Pauli matrices form a complete basis for the space of \(2^n \times 2^n\) Hermitian matrices, meaning any Hermitian matrix can be expressed as their real linear combination.
The proof is divided into the following five steps:\\
\\
The first step involves basis construction. For \(n\) qubits, the Hilbert space dimension is \(2^n\), and the matrix dimension is \(2^n \times 2^n\). The set of tensor products of Pauli matrices is defined as:
\begin{align}
    \{P_{k_1} \otimes P_{k_2} \otimes \cdots \otimes P_{k_n} \mid P_{k_i} \in \{I, \sigma_x, \sigma_y, \sigma_z\}\}
\end{align}
Each position has 4 choices (\(I, \sigma_x, \sigma_y, \sigma_z\)), and there are \(n\) positions, so there are a total of:
$
4^n = (2^2)^n
$
tensor product matrices.\\
\\
Step 2 is to prove linear independence. These tensor product matrices are linearly independent, which can be proven through their trace orthogonality. For two different tensor products \(P_{\mathbf{k}} = P_{k_1} \otimes \cdots \otimes P_{k_n}\) and \(P_{\mathbf{m}} = P_{m_1} \otimes \cdots \otimes P_{m_n}\):
\begin{align}
\text{tr}(P_{\mathbf{k}}^{\dagger} P_{\mathbf{m}}) = \text{tr}(P_{k_1}^{\dagger} P_{m_1}) \cdot \text{tr}(P_{k_2}^{\dagger} P_{m_2}) \cdots \text{tr}(P_{k_n}^{\dagger} P_{m_n})
\end{align}
The trace properties of single Pauli matrices are:\\
\(\text{tr}(I) = 2\), \(\text{tr}(\sigma_x) = \text{tr}(\sigma_y) = \text{tr}(\sigma_z) = 0\),
 \(\text{tr}(I \cdot I) = 2\), \(\text{tr}(\sigma_i \cdot \sigma_j) = 2\delta_{ij}\) for \(i, j = x, y, z\), \(\text{tr}(I \cdot \sigma_i) = 0\)

Therefore:
\begin{align}
\text{tr}(P_{\mathbf{k}}^{\dagger} P_{\mathbf{m}}) = 2^n \delta_{\mathbf{k}\mathbf{m}}
\end{align}
That is, the trace is \(2^n\) only when \(\mathbf{k} = \mathbf{m}\) (all \(k_i = m_i\)); otherwise, it is 0. This proves linear independence.\\
\\
The third step is dedicated to spanning the entire matrix space. The space of \(2^n \times 2^n\) complex matrices is characterized by a dimension of \( (2^n)^2 = 4^n \), a result derived from the fact that each entry in such a matrix can be an independent complex number, leading to a total of \(2^n \times 2^n\) free parameters that define the matrix. Crucially, the set of tensor product matrices under consideration contains exactly \(4^n\) linearly independent elements, a count that directly matches the dimension of the matrix space. This correspondence is not coincidental: in linear algebra, a set of linearly independent vectors (or matrices, viewed as vectors in a higher-dimensional space) spans the entire space if their number equals the space's dimension. Thus, the tensor product matrices, by virtue of having \(4^n\) linearly independent elements, are sufficient to generate every possible \(2^n \times 2^n\) complex matrix through linear combinations. n leads to exponentially large matrix spaces that require structured bases for tractable manipulation.\\
\\
Next step we use the hermiticity constraint. Each tensor product \( P_{\mathbf{k}} = P_{k_1} \otimes \cdots \otimes P_{k_n} \) satisfies the Hermiticity constraint, meaning \( P_{\mathbf{k}}^{\dagger} = P_{\mathbf{k}} \). This property arises from the fundamental Hermiticity of its constituent factors: each \( P_{k_i} \) is either the identity matrix \( I \) or one of the Pauli matrices, all of which are Hermitian (i.e., \( P_{k_i}^{\dagger} = P_{k_i} \)). For tensor products, the adjoint operation distributes across the tensor product such that \( (P_{k_1} \otimes \cdots \otimes P_{k_n})^{\dagger} = P_{k_1}^{\dagger} \otimes \cdots \otimes P_{k_n}^{\dagger} \); substituting the Hermiticity of each factor confirms that the entire tensor product matrix is Hermitian. Turning to the space of \( 2^n \times 2^n \) Hermitian matrices, its dimension can be derived by analyzing the degrees of freedom in such matrices: diagonal elements must be real numbers (contributing \( 2^n \) independent parameters), while off-diagonal elements exhibit conjugate symmetry (i.e., \( M_{ij} = \overline{M_{ji}} \) for \( i \neq j \)), reducing the independent parameters to \( \frac{2^n(2^n - 1)}{2} \) complex numbers (each contributing 2 real degrees of freedom). Summing these contributions gives a total of \( 2^n + 2 \cdot \frac{2^n(2^n - 1)}{2} = (2^n)^2 = 4^n \) degrees of freedom, matching the dimension of the full space of \( 2^n \times 2^n \) complex matrices. \\
\\
Then comes the final step. Given that the basis matrices \( P_{\mathbf{k}} \) are Hermitian, any \( 2^n \times 2^n \) Hermitian matrix \( A \) can be expressed as a linear combination of these basis matrices, specifically in the form \( A = \sum_{\mathbf{k}} c_{\mathbf{k}} P_{\mathbf{k}} \), where \( c_{\mathbf{k}} \) denote the coefficients of the combination. To ensure the Hermiticity of \( A \), we examine its adjoint: \( A^{\dagger} = \sum_{\mathbf{k}} c_{\mathbf{k}}^* P_{\mathbf{k}}^{\dagger} \). Since each \( P_{\mathbf{k}} \) is Hermitian, \( P_{\mathbf{k}}^{\dagger} = P_{\mathbf{k}} \), simplifying the expression for \( A^{\dagger} \) to \( \sum_{\mathbf{k}} c_{\mathbf{k}}^* P_{\mathbf{k}} \). For \( A \) to be Hermitian, it must satisfy \( A = A^{\dagger} \), which implies \( \sum_{\mathbf{k}} c_{\mathbf{k}} P_{\mathbf{k}} = \sum_{\mathbf{k}} c_{\mathbf{k}}^* P_{\mathbf{k}} \). Given the linear independence of the basis matrices \( P_{\mathbf{k}} \), the coefficients of corresponding terms in the sums must be equal, leading to the condition \( c_{\mathbf{k}} = c_{\mathbf{k}}^* \) for all \( \mathbf{k} \). This equality indicates that each coefficient \( c_{\mathbf{k}} \) must be a real number, i.e., \( c_{\mathbf{k}} \in \mathbb{R} \). \\
\\
To summarize, The \(4^n\) Hermitian and linearly independent Pauli tensor product matrices span the entire space of \(2^n \times 2^n\) Hermitian matrices. Therefore, any \(2^n \times 2^n\) Hermitian matrix \(A\) can be decomposed as:
\begin{align}
A = \sum_{\mathbf{k}} c_{\mathbf{k}} P_{k_1} \otimes \cdots \otimes P_{k_n}
\end{align}
where \(c_{\mathbf{k}}\) are real numbers.\\
\\
Finally, we present the specific method for decomposition. Given a \(2^n \times 2^n\) Hermitian matrix \(A\), we can compute its decomposition coefficients using trace orthogonality.\\
\\
The \textbf{first step} is coefficient calculation. The specific steps are as follows:
\begin{itemize}
    \item For each basis matrix \(P_{\mathbf{k}} = P_{k_1} \otimes \cdots \otimes P_{k_n}\), the coefficient \(c_{\mathbf{k}}\) is:
\begin{align}
        c_{\mathbf{k}} = \frac{1}{2^n} \text{tr}(A \cdot P_{\mathbf{k}})
\end{align}
    \item Reason:
\begin{align}
        \text{tr}(P_{\mathbf{k}}^{\dagger} P_{\mathbf{m}}) = 2^n \delta_{\mathbf{k}\mathbf{m}}
\end{align}
        If \(A = \sum_{\mathbf{m}} c_{\mathbf{m}} P_{\mathbf{m}}\), then:
\begin{align}
        \text{tr}(A \cdot P_{\mathbf{k}}) = \text{tr}\left(\sum_{\mathbf{m}} c_{\mathbf{m}} P_{\mathbf{m}} \cdot P_{\mathbf{k}}\right) = \sum_{\mathbf{m}} c_{\mathbf{m}} \text{tr}(P_{\mathbf{m}} \cdot P_{\mathbf{k}}) = c_{\mathbf{k}} \cdot 2^n
\end{align}
  \begin{align}
        c_{\mathbf{k}} = \frac{1}{2^n} \text{tr}(A \cdot P_{\mathbf{k}})
  \end{align}
        \(\text{tr}(A \cdot P_{\mathbf{k}})\) is real (since both \(A\) and \(P_{\mathbf{k}}\) are Hermitian), ensuring \(c_{\mathbf{k}} \in \mathbb{R}\).
\end{itemize}
\vspace{0.3cm}
The \textbf{second step} is constructing the decomposition. We only need to substitute all coefficients:
      \begin{align}
        A = \sum_{\mathbf{k}} \left(\frac{1}{2^n} \text{tr}(A \cdot P_{\mathbf{k}})\right) P_{\mathbf{k}}
     \end{align}
     Iterate over all \(\mathbf{k} \in \{0, 1, 2, 3\}^n\) (a total of \(4^n\) terms) to obtain the complete decomposition.

\end{document}